\documentclass[12pt,preprint]{aastex}
\newcommand{\spitzer}{{\it Spitzer}}
\newcommand{\htwo}{H\,{\sc ii}}
\usepackage{natbib}
\usepackage{epsfig}

\begin{document}

\title{Disk Evolution in W5: Intermediate Mass Stars at 2--5~Myr}

\author{Xavier P. Koenig,\altaffilmark{1} Lori E. Allen\altaffilmark{2}}
\altaffiltext{1}{Harvard-Smithsonian Center for Astrophysics, 60 Garden Street, 
Cambridge, MA, USA
Current address: NASA GSFC, 8800 Greenbelt Road, Greenbelt, MD 20771}
\altaffiltext{2}{NOAO, 950 North Cherry Avenue, Tucson, AZ, USA}

\begin{abstract}
  We present the results of a survey of young intermediate mass stars
  (age $<$~5~Myr, 1.5 $<M_{\star} \leq $ 15~$M_{\odot}$) in the W5
  massive star forming region. We use combined optical, near-infrared
  and {\it Spitzer} Space Telescope photometry and optical
  spectroscopy to define a sample of stars of spectral type A and B
  and examine their infrared excess properties. We find objects with
  infrared excesses characteristic of optically thick disks,
  i.e. Herbig AeBe stars. These stars are rare: $<$1.5\% of the entire
  spectroscopic sample of A and B stars, and absent among stars more
  massive than 2.4~$M_\odot$. 7.5\% of the A and B stars possess
  infrared excesses in a variety of morphologies that suggest their
  disks are in some transitional phase between an initial, optically
  thick accretion state and later evolutionary states. We identify
  four morphological classes based on the wavelength dependence of the
  observed excess emission above theoretical photospheric levels: (a)
  the optically thick disks; (b) disks with an optically thin excess
  over the wavelength range 2 to 24~$\micron$, similar to that shown
  by Classical Be stars; (c) disks that are optically thin in their
  inner regions based on their infrared excess at 2--8~$\micron$ and
  optically thick in their outer regions based on the magnitude of the
  observed excess emission at 24~$\micron$; (d) disks that exhibit
  empty inner regions (no excess emission at $\lambda$ $\leq$
  8~$\micron$) and some measurable excess emission at 24~$\micron$. A
  sub-class of disks exhibit no significant excess emission at
  $\lambda \leq$ 5.8~$\micron$, have excess emission only in the {\it
    Spitzer} 8~$\micron$ band and no detection at 24~$\micron$. We
  discuss these spectral energy distribution (SED) types, suggest
  physical models for disks exhibiting these emission patterns and
  additional observations to test these theories.
\end{abstract}

\keywords{accretion, accretion disks --- circumstellar matter --- infrared: stars --- planetary systems: protoplanetary disks --- stars: pre-main sequence --- techniques: spectroscopic}

\section{Introduction}
Recent observations have shown that planets exist around intermediate
mass stars (spectral types A and B, 1.5 $<M_{\star} \leq $
15~$M_{\odot}$, Rivera et~al. 2005, Johnson et~al. 2007) as well as
around solar type ($<M_{\star} \sim $ 1~$M_{\odot}$) stars. Extensive
near-infrared (IR), mid-infrared (IR) through submillimeter and
millimeter observations of low mass star disks (stellar spectral type
later than $\sim$K5) have shown that $\sim$90\% of these stars lose
their optically thick inner disks by an age of 5--7~Myr
\citep{strom89, haisch01, hillen08}. Early studies of the more massive
Herbig Ae/Be stars (M$_{star}$ = 2--8~M$_{\odot}$) indicate that their
disks may be dissipated within a shorter timescale $\sim$3~Myr
\citep{hern05}. Such surveys however, have been constrained to small
samples of intermediate mass stars in several nearby star forming
regions. In order to better compare the low mass star studies with a
large sample of intermediate mass stars in a single region we have
embarked on a combined infrared and optical survey of the massive star
forming region W5 \citep{wester58,koenig08}.

Combined optical, near-IR ($JHK_S$) and {\it Spitzer} spectral energy
distribution (SED) analyses provide us with a simple diagnostic of the
dust in young stellar disks. We have conducted a study of spectral
types and H$\alpha$ equivalent widths from optical spectra, combined
with a simple classification scheme for infrared disk SED morphologies
of A and B stars in W5. $\S$~\ref{sec:obs} describes our
observations. In $\S$~\ref{sec:colors} we present our analysis of the
different types of disks we detect in the sample of B and A stars in
W5, and our procedure for filtering foreground and background
contamination. In $\S$~\ref{sec:discuss} we discuss our results for
infrared disk types in the context of models of young stellar disk
evolution.
 
\section{Data}\label{sec:obs}
\subsection{Spitzer Photometry}
Our {\it Spitzer} IRAC and MIPS observations of W5 are described in
detail in \citet{koenig08}. The IRAC \citep{fazio04} observations (PID
20300) were broken down into three rectangular Astronomical Observing
Requests (AORs) covering $\sim$1.8{\degr} $\times$ 1.6{\degr}, in
order to observe at multiple rotation angles and help minimize
artifacts aligned along columns or rows of the array. Each AOR had a
coverage of 1 High Dynamic Range (HDR) frame. HDR mode results in a
10.4~s and 0.4~s exposure being taken at each position in each map. We
used the $\texttt{clustergrinder}$ software tools developed by
R. Gutermuth to produce final image mosaics from these data in each
wavelength band. $\texttt{Clustergrinder}$ incorporates all necessary
image treatment steps, for example, saturated pixel processing and
distortion corrections \citep[see][for a more complete description of
the processing performed]{gutermu08}. $\texttt{Clustergrinder}$ uses
the short 0.4~s exposures only in saturated or near-saturated regions,
so that the combined map has an effective total integration time of
3$\times$10.4 s = 31.2~s in most of the overlapping areas. We also
incorporated in our data processing archival data covering AFGL 4029
\citep[from {\it Spitzer} GTO program PID 201,][]{allen05}, at the
eastern end of W5.

The MIPS \citep{rieke04} observations were carried out on 2006
February 23 UT under our GO-2 program, PID 20300. Images were taken in
scan-map mode using the medium scan speed for an average exposure time
of 41.9~s pixel$^{-1}$ once frames were combined. The raw data were
processed with pipeline version S13.2.0. We produced final mosaics
using the MIPS instrument team Data Analysis Tool, which calibrates
the data and applies a distortion correction to each individual
exposure before combining \citep{gordon05}. We used only the 24
$\micron$ band data for our analysis in this paper, since strong
background emission dominates at the longer wavelength (70 and 160
$\micron$) bands of MIPS, and lower sensitivity reduces the number of
detectable objects to a level not useful for the present study.

We carried out point source extraction and aperture photometry of all
point sources on the final IRAC mosaics with PhotVis version
1.10beta3. PhotVis is an IDL GUI-based photometry visualization tool
\citep[see][]{gutermu04} that utilizes DAOPHOT modules ported to IDL
in the IDL Astronomy User's Library \citep{landsman93}. We used
PhotVis to visually inspect the detected sources in IRAC bands 1, 3
and 4, adding sources not detected automatically, but clearly visible
in the images with the GUI tool and rejected any structured nebulosity
or cosmic rays mistaken for stellar sources by the automatic detection
algorithm. To save time, we did not visually check the band 2
photometry in this manner. Instead, we took the cleaned band 1 source
list as the starting point for finding objects in the image and
extracted photometry at each position. Radii of the apertures and
inner and outer limits of the sky annuli were 2.4{\arcsec},
2.4{\arcsec} and 7.2{\arcsec} respectively. The photometry was
calibrated using large-aperture in-flight measurements of standard
stars, with an appropriate aperture correction in each channel to
correct for the smaller apertures used in this study.

Averaged over the whole W5 field, our source catalog is 90\% complete
to a magnitude of 15.5 at 3.6 $\micron$, 15.5 at 4.5 $\micron$, 14.0
at 5.8 $\micron$ and 12.7 at 8 $\micron$. The completeness is less in
regions of bright diffuse emission: $\sim$14 at 3.6 and 4.5 $\micron$,
$\sim$11 at 5.8 $\micron$ and $\sim$9.5 at 8.0 $\micron$
\citep{koenig08}.

We conducted point source extraction and aperture photometry of point
sources in the 24~$\micron$ MIPS mosaic using the
point-spread-function fitting capability in IRAF DAOPHOT
\citep{stetson87}. We visually inspected the image to pick out point
sources not automatically detected due to bright diffuse emission
evident throughout the image. We match the four-band IRAC source list
to the MIPS catalog using a 2{\arcsec} search radius, selecting the
object closest to the MIPS point spread function centroid in cases
where more than one IRAC object is a match.

\subsection{Optical Photometry}
We used KeplerCam \citep{sze05} on the 1.2~m telescope at FLWO to
image six fields in W5 in Sloan $r$ and $z$ filters on 2006 January
21. KeplerCam has a monolithic 4096$^2$ CCD detector giving a
23.1{\arcmin}$\times$23.1{\arcmin} field of view. We binned the images
2 $\times$ 2, giving a scale of 0.68{\arcsec} per binned pixel. Each
field was imaged with a series of 10~s and 180~s exposures to detect
both faint and bright objects. Sky subtraction was carried out by
constructing sky images from median-combined data frames. We extracted
simple aperture photometry with IRAF DAOPHOT $\texttt{daofind}$ and
$\texttt{phot}$ tasks.

We observed a further 6 fields in W5 on 2006 September 25 with MegaCam
\citep{mcle00} on the MMT to fill in gaps left by the KeplerCam
pointings. MMT/MegaCam has 36 chips with 2048$\times$4608 pixels, a
pixel scale of 0.08{\arcsec} pixel$^{-1}$ and a total field of view
(FOV) of 24{\arcmin}. We obtained three 0.4~s and three 50~s dithered
exposures in the MegaCam $r'$ filter, and three 0.4~s and three 70~s
dithered exposures in the MegaCam $z'$ filter in gray conditions, with
0.8{\arcsec}--2.0{\arcsec} image quality in the $r'$ images and
0.96{\arcsec}--1.8{\arcsec} image quality in the $z'$ images. We
reduced the data based on the method described in Matt Ashby's MegaCam
Reduction Guide. Our reduction relied in part on software written
specifically for MMT/MegaCam data reduction by Brian McLeod. We used
the Two Micron All Sky Survey \citep[2MASS][]{skrutskie06} stellar
catalog to derive precise astrometric solutions for each science
exposure. We used images of the SSA 22 field \citep{lilly91} taken on
the same night to construct sky frames for proper sky and fringe
subtraction. We also used dithered images of the SSA 22 field to
derive illumination correction images in $z'$ and $r'$ to divide out
the variation in zero point across MegaCam’s FOV. We did a weighted
coaddition of the reduced images using the IRAF $\texttt{imcombine}$
task. We extracted photometry in $r'$ and $z'$ filters with the
PhotVis tool as described above.

\subsection{Optical Spectra}
\subsubsection{Hectospec/6.5m}
We obtained optical spectra of candidate intermediate mass stars with
the Hectospec multifiber spectrograph mounted on the 6.5~m MMT
telescope on Mount Hopkins \citep{fab94}. Hectospec is a multi-object
spectrograph with 300 fibers that can be placed within a 1{\degr}
diameter circular field. We used the 270 groove mm$^{-1}$ grating, and
obtained spectra in the range 3700--9000 {\AA} with a resolution of
6.2~{\AA}. The pointing is fixed using 2 to 3 guide stars. Hectospec
requires better than 0{\farcs}3 coordinates for fiber positioning; the
fibers subtend a 1{\farcs}5 diameter circle. We used guide stars drawn
from 2MASS; our {\it Spitzer} and optical photometry coordinates are
also registered to 2MASS. Hectospec allowed us to observe a large
sample in a series of queue observing runs from October 2006 to
December 2008.

Data reduction was performed by S. Tokarz through the CfA Telescope
Data Center, using IRAF tasks and other customized reduction
scripts. The reduction procedure was the standard for Hectospec data,
with the addition of our special sky subtraction procedure. In order
to avoid the difficulties encountered when subtracting a sky spectrum
made from an average over the highly variable H\,{\sc ii} region, we
took exposures offset by $\sim$5{\arcsec} after each one of the
configurations. In this way we obtained a sky spectrum very close to
each star through the same fiber. Background subtraction was performed
in IDL to remove each wavelength calibrated sky offset spectrum from
its corresponding wavelength calibrated object spectrum. Since each
object spectrum was made up of three averaged observations and the sky
spectra were only single exposures, a scale factor was calculated at
5600{\AA} and 8400{\AA} to completely remove the sky lines at these
points. These two points were used to linearly interpolate a scale
factor over the whole wavelength range to correct each sky offset
spectrum.

\subsubsection{FAST/1.5m}
\begin{figure} 
\centering
\epsfig{file=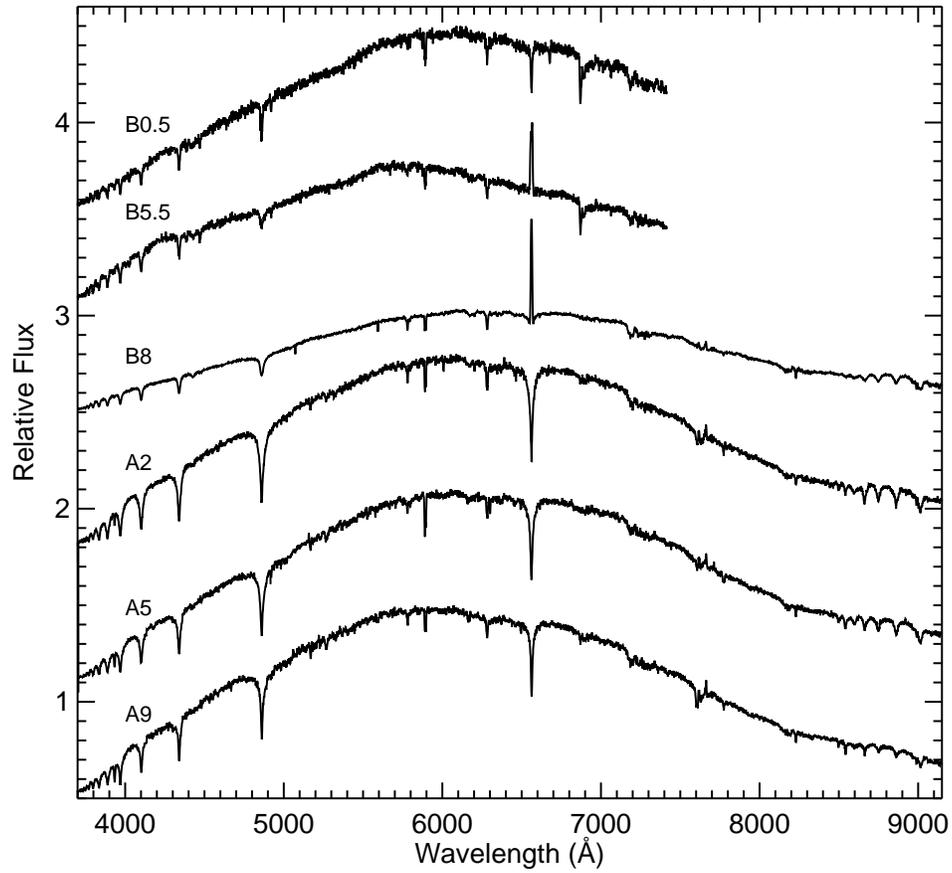,width=0.8\linewidth,clip=}
\caption{Sample FAST (shorter wavelength scale) and sky-subtracted
  Hectospec spectra.\label{fig:sampspec} Relative flux (offset by an
  arbitrary factor) is plotted against wavelength. Spectral types
  assigned by our automated code are given alongside each object.}
\end{figure}

Additional spectra were obtained using the FAST slit spectrograph
mounted on the 1.5~m telescope at FLWO \citep{fab94fast,fab98}, using
a Loral 512$\times$2688 CCD. We used the standard FAST COMBO
configuration, with a 300 groove mm$^{-1}$ grating and a 3{\arcsec}
wide slit. This setup gave spectra in the range 3800--7200 {\AA}, with
a resolution of $\sim$6 {\AA}, comparable to the Hectospec
spectra. The stars were observed during several queued runs from 2007
September until 2008 December (observers: P. Berlind \&
M. Calkins). The spectra were reduced at the CfA (S. Tokarz) using
software developed specifically for the FAST COMBO observations
\citep{tokarz97}. A sample of spectra is displayed in
Figure~\ref{fig:sampspec}.

\subsection{Spectral Classification}\label{sec-spclass}
We classified our objects using the spectral classification software
SPTCLASS\footnote{http://www.astro.lsa.umich.edu/$\sim$hernandj/SPTclass/sptclass.html}
\citep{hern04}. SPTCLASS works well over the FAST and Hectospec
wavelength ranges for stars with spectral types in the range from
$\sim$O9 to $\sim$L0. The main code includes three spectral
classification schemes: the first is optimized to classify stars in
the mass range of T Tauri stars (TTS, type K5 or later), the second is
optimized to classify stars in the mass range of solar-type stars
(late F to early K), and the third is optimized to classify stars in
the mass range of HAeBe stars (F5 or earlier). The schemes are based
on 66 spectral indices sensitive to changes in T$_{eff}$ but
insensitive to reddening, stellar rotation, luminosity class, and
S/N. The code generates a postscript file which shows the individual
results from the different spectral indices, and a closer view around
the H$\alpha$ and Li I (6707{\AA}) lines. The best result from the
three schemes was selected for each spectrum by visual inspection---in
some cases (about 5--10\% of stars) where the code failed to
accurately classify a spectrum, a type was assigned by visual
comparison to standards in the literature
\citep{jacob84,andrill,carq}.

\section{Analysis}

\subsection{A and B Star Sample Selection}
\begin{figure}
\centering
\epsfig{file=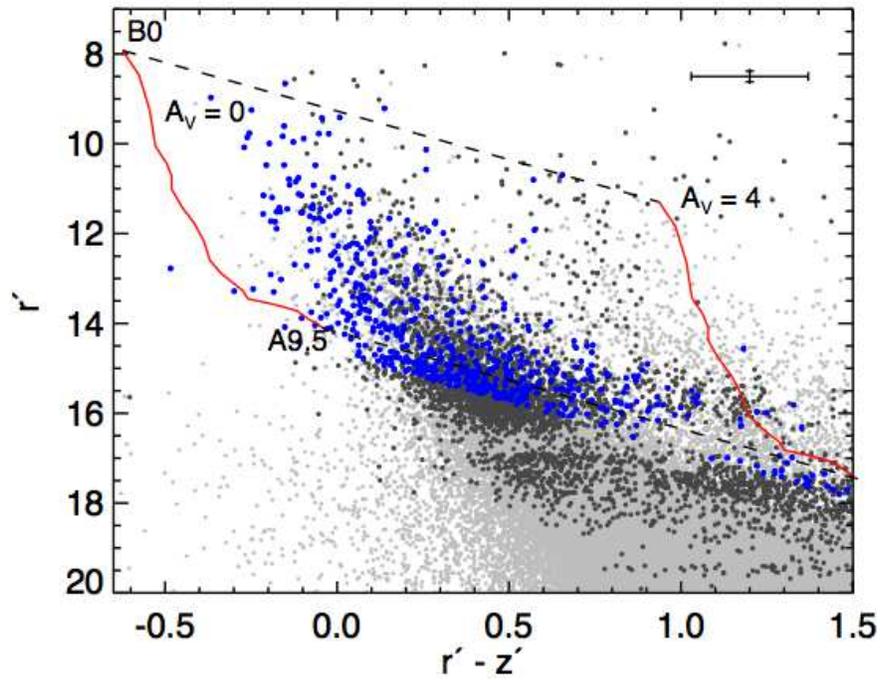,width=0.7\linewidth,clip=}
\caption{W5 optical photometric sample (light gray points) overlaid
  with dark gray points showing spectral sample.~\label{fig:extsamp}
  Blue points mark A and B stars in the spectral catalog that fall
  within the extinction limit. A typical error bar is shown in upper
  right.}
\end{figure}

In this paper we specifically target stars of spectral type A and B,
roughly corresponding to masses 1.5 $<M_{\star} \leq $
15~$M_{\odot}$. We select an extinction-limited sample of A and B
stars from our spectral catalog. Figure~\ref{fig:extsamp} shows the
observed color-magnitude diagram for the optical photometric catalog
in W5 (light gray points), overlaid with the photometry for the
complete spectral catalog in dark gray. We show a theoretical main
sequence in $r'$ and $r'-z'$ (in red) derived from the absolute
magnitude scale given in \citet{schm82} and the optical colors from
\citet{kenyon95} converted to the Sloan filter set using the
color-transform relations given in \citet{jordi06}. The main sequence
is shown as it would appear at a distance of 2~kpc and extinction
$A_V$=0 and $A_V$=4. We highlight in blue objects from the spectral
catalog with spectral type A or B whose photometry $\pm$1$\sigma$
brings them within the trapezoidal region marked in the figure. This
population of objects makes up our sample: a total of 610
spectroscopically confirmed A and B stars.

\subsection{The Age of W5}\label{sec:age} 

The full spectroscopic sample in W5 comprises 4800 stars, ranging in
spectral type from O9 to M6.5. Since we don't have a good technique to
establish membership of the region, we select objects which match the
\spitzer\ excess source list from \citet{koenig08}: a total of 389
stars. We calculate for each star the value of optical extinction
$A_V$ from the observed optical photometry:

\begin{equation}
A_V = 2.56\left( r' - z' - \left[1.584(R_C - I_C) - 0.386\right] \right)\label{eq:extinc}
\end{equation}

We obtain $R_C-I_C$ from \citet{kenyon95} according to spectral
type. Equation \ref{eq:extinc} relies on the conversion between
$R_CI_C$ and SDSS\footnote{http://www.sdss.org} $r'z'$ given in
\citet{jordi06} and the extinction versus wavelength values given in
\citet{schlegel98} such that $A_{r'}$ = 0.843~$A_V$ and $A_{z'}$ =
0.453~$A_V$.

\begin{figure}
\centering
\begin{tabular}{cc}
\epsfig{file=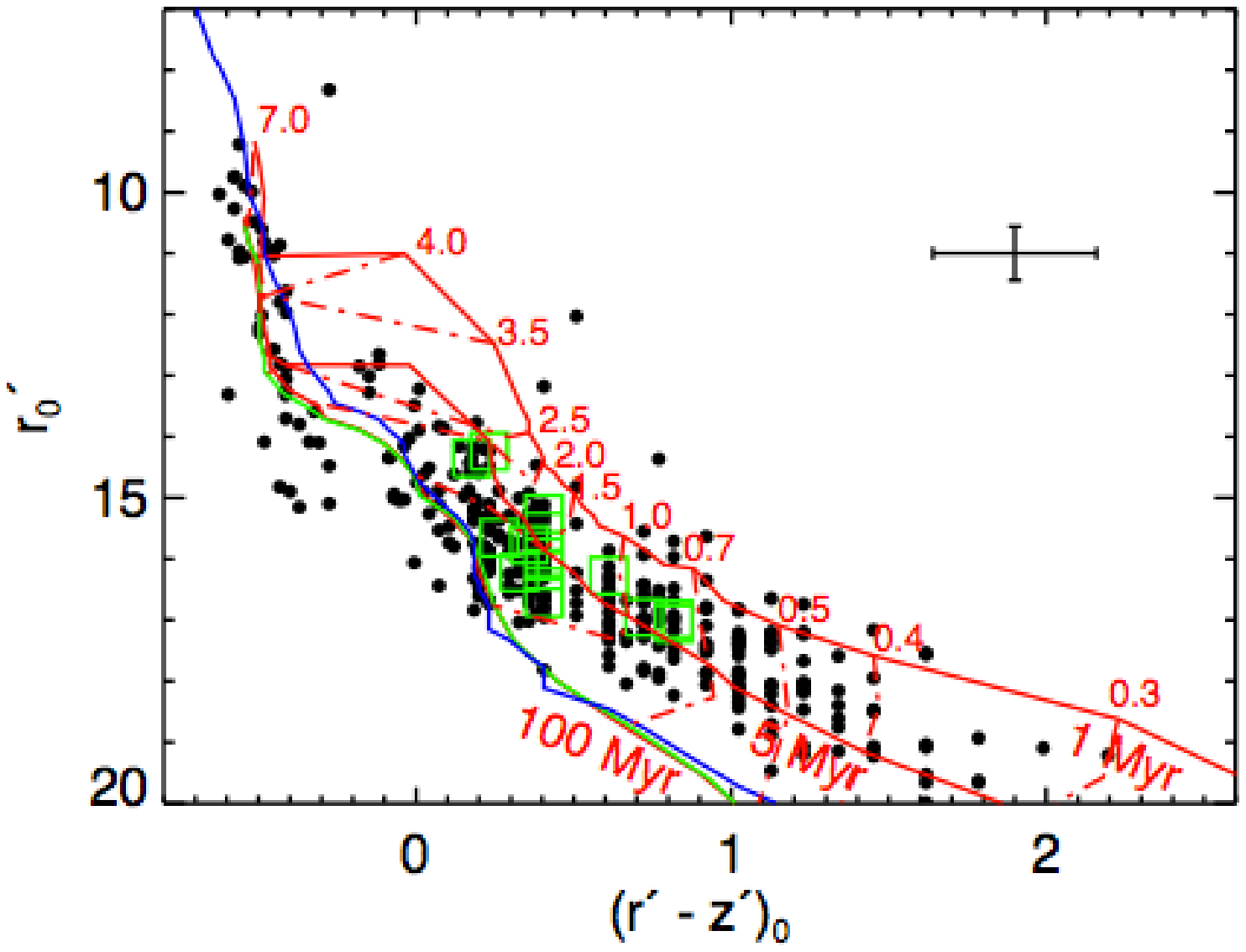,width=0.45\linewidth,clip=} &
\epsfig{file=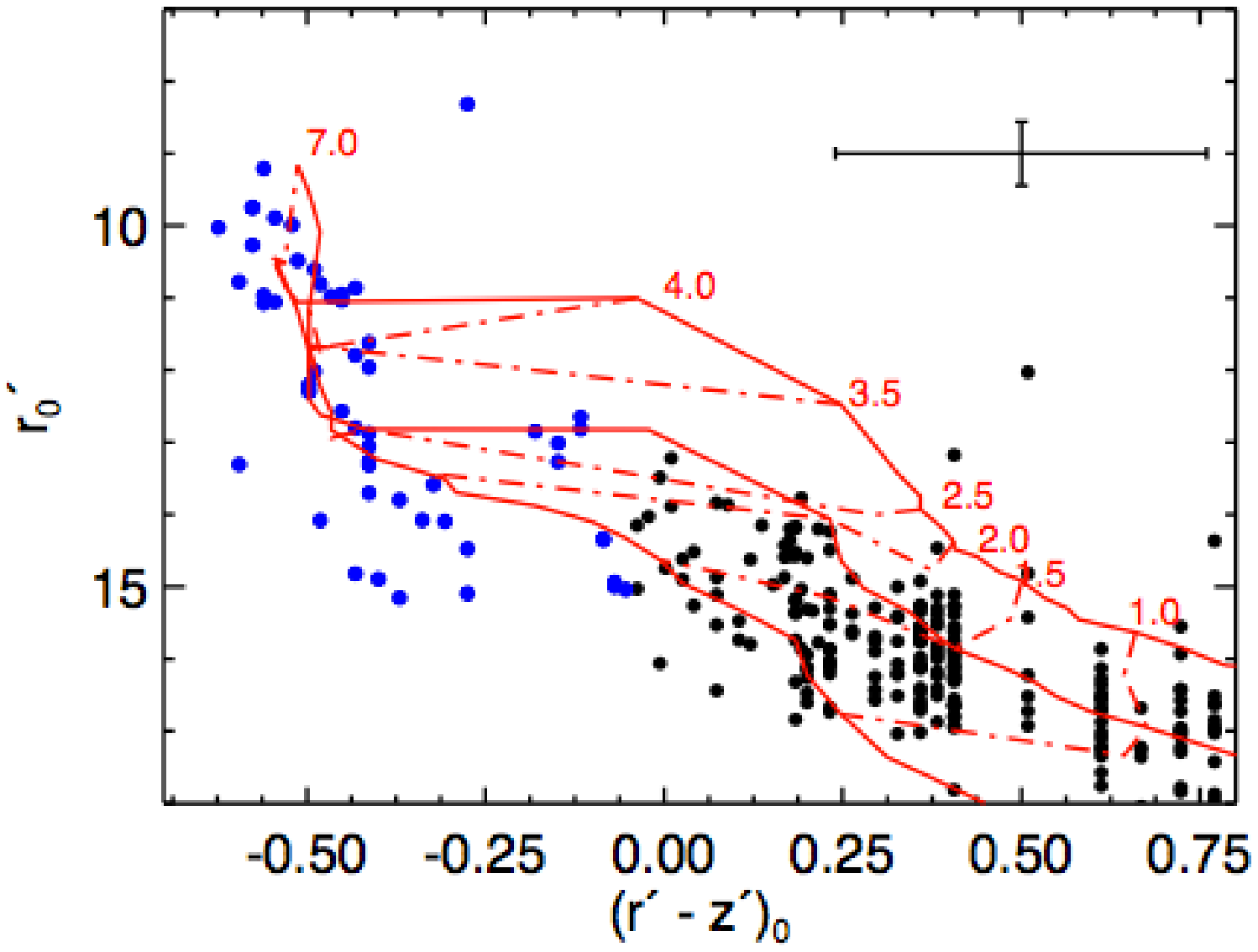,width=0.45\linewidth,clip=} 
\end{tabular}
\caption{Left panel: dereddened color-magnitude diagram for all
  \spitzer\ excess sources (Koenig et~al. 2008) detected in our
  spectroscopic survey.\label{fig:age} Green boxes indicate those
  objects with convincing Lithium absorption. Solid red lines:
  isochrones at 1, 5 and 100~Myr from \citet{siess00}. Solid blue
  line: Main Sequence from \citet{kenyon95} and \citet{schm82}; solid
  green line: Main Sequence from \citet{siess00}. Right panel:
  zoomed-in portion showing the A and B stars highlighted with blue
  points. A typical error bar is shown in upper right of each panel.}
\end{figure}

In Figure~\ref{fig:age} (left panel) we present the dereddened optical
color-magnitude diagram for this sub-sample. We overplot isochrones at
1, 5 and 100~Myr from \citet{siess00}, assuming a distance of 2~kpc to
W5 and using the conversion relations between $R_CI_C$ and SDSS $r'z'$
from \citet{jordi06}. The distribution of the later spectral type
stars at $(r'-z')_0>0.3$ suggests that W5 is at age 5~Myr or
younger. This compares well with the age upper limit implied by the
presence of the central O stars in the region \citep[5~Myr,][]{karr03}
and with the bubble expansion ages presented in \citet{vallee79} (age
$<$ 1.4~Myr). However, a trend of increasing stellar age with
increasing mass is apparent in Fig.~\ref{fig:age}. In the right panel
of the figure we show a zoomed-in portion of the diagram focusing on
the massive stars, with A and B stars highlighted in blue. In this
part of the color-magnitude diagram the stars lie fainter than where
we would expect them at an age of 5~Myr given the models of
\citet{siess00}. Some small fraction of these objects may be
background stars, but some are likely intrinsically fainter than the
model prediction. This result is a common property of stellar ages
derived using the HR diagram as noted by \citet{hillen08age}. We
believe this does not indicate a true difference in the ages of the
low and high mass stars in W5.

The spectral sample was of sufficient resolution to detect Lithium
$\lambda$6707 absorption, an indicator of youth \citep{rand01,hart03}
in 15 of these objects. We highlight these stars with green boxes in
Figure~\ref{fig:age}.

We present the photometry and derived spectral types, H$\alpha$
equivalent widths (EW) and $A_V$ extinction values for the list of 389
IR excess stars with spectral types described above in
Table~\ref{tab-spectra}. Column 14 gives the spectral type and its
respective error for each star. The uncertainty is a combination of
error from the fit of each index to the standard main sequence and the
error in the measurement of each index. We measured the equivalent
width of the H$\alpha$ line using the $\texttt{splot}$ task in the
IRAF spectral reduction package $\texttt{noao.onedspec}$. In those
cases where both emission and absorption components were visible in
the spectrum, the EW was measured across the full line profile. The
EW(H$\alpha$) is given in column 15 of Tab.~\ref{tab-spectra}. We list
the extinction $A_V$ in column 16.

\begin{deluxetable}{lccccccccccccccc}
\tabletypesize{\tiny}
\rotate
\tablewidth{0pt}
\tablecaption{W5 IR Excess Spectral Catalog}
\tablehead{\colhead{ } & \colhead{RA} & \colhead{Dec} & \colhead{$J$} & \colhead{$H$} & \colhead{$K_S$} & \colhead{$[3.6]$} & \colhead{$[4.5]$} & \colhead{$[5.8]$} & \colhead{$[8.0]$} & \colhead{$[24]$} & \colhead{$r'$} & \colhead{$z'$} & \colhead{Sp. Type} & \colhead{EW(H$\alpha$)} & \colhead{$A_V$} \\
\colhead{ID\tablenotemark{a}} & \colhead{(deg)} & \colhead{(deg)} & \colhead{(mag)} & \colhead{(mag)} & \colhead{(mag)} & \colhead{(mag)} & \colhead{(mag)} & \colhead{(mag)} & \colhead{(mag)} & \colhead{(mag)} & \colhead{(mag)} & \colhead{(mag)} & \colhead{Type\tablenotemark{b}} & \colhead{($\textrm{\AA}$)} & \colhead{(mag)}}
\startdata
1141 & 41.831320 & 60.498171 & 14.11(04) & 13.38(03) & 13.01(03) & 12.52(01) & 12.34(01) & 10.35(01) & 8.41(01) & 5.98(15) & 17.06(12) & 15.89(12) & A8.0$\pm$1.0 & 8.12 & 3.22(50) \\
1473 & 41.922683 & 60.521748 & 14.05(04) & 13.13(04) & 12.30(03) & 11.13(00) & 10.78(00) & 10.47(01) & 9.91(01) & 5.76(05) & 17.21(12) & 16.06(12) & F6.0$\pm$1.0 & 0.32 & 2.77(50) \\
1562 & 41.952386 & 60.964075 & 12.04(02) & 10.94(02) & 9.96(02) & 8.85(00) & 8.55(00) & 8.08(00) & 7.11(00) & 3.59(07) & 14.06(12) & 13.50(12) & B9.0$\pm$0.5 & -46.50 & 2.48(50) \\
1576 & 41.958330 & 60.679266 & 14.64(03) & 13.93(04) & 13.59(04) & 13.41(01) & 13.07(01) & 13.38(07) & \nodata & \nodata & 18.17(12) & 16.52(13) & A1.0$\pm$0.5 & 4.31 & 5.19(52) \\
1675 & 41.986977 & 60.502611 & 16.26(11) & 14.93(09) & 14.22(06) & 13.22(02) & 12.87(01) & 12.46(04) & 11.85(08) & 8.87(16) & 18.58(15) & 17.84(13) & M7.0$\pm$1.0 & -26.38 & \nodata \\
1848 & 42.036912 & 60.338789 & 14.45(03) & 13.40(03) & 13.08(03) & 12.79(01) & 12.75(01) & 12.43(05) & 11.76(12) & \nodata & 18.50(12) & 16.61(12) & K3.5$\pm$1.0 & 2.27 & 3.82(51) \\
1970 & 42.078841 & 60.337579 & 13.56(02) & 13.12(02) & 12.93(02) & 12.78(01) & 12.72(01) & 12.40(09) & 11.52(14) & \nodata & 15.90(12) & 15.12(12) & B9.0$\pm$0.5 & 9.35 & 3.06(50) \\
2068 & 42.115392 & 60.711151 & 13.92(03) & 12.83(03) & 12.05(02) & 10.70(00) & 10.02(00) & 9.11(00) & 7.58(01) & 4.46(03) & 17.33(12) & 15.96(12) & F7.5$\pm$1.0 & -27.17 & 3.22(50) \\
2234 & 42.175010 & 60.686677 & 14.04(03) & 13.63(04) & 13.49(04) & 13.48(01) & 13.45(01) & 13.11(07) & 12.77(20) & \nodata & 15.71(12) & 15.27(12) & F3.5$\pm$0.5 & 3.77 & 1.15(50) \\
2309 & 42.204412 & 60.792946 & 13.52(03) & 12.55(03) & 11.76(02) & 10.31(00) & 9.93(00) & 9.60(01) & 8.96(01) & 4.92(06) & 17.22(12) & 15.61(12) & G7.0$\pm$1.0 & -2.50 & 3.58(50)
\enddata
\tablecomments{This Table is published in its entirety in the electronic edition of the {\it Astrophysical Journal}.\label{tab-spectra} A portion is shown here for guidance regarding its form and content. Values in parentheses by photometry signify error in last two digits of magnitude value. Right Ascension and Declination coordinates are J2000.0. Please contact author for full table.}
\tablenotetext{a}{ID numbering is the same as that from Koenig et~al. (2008).}
\tablenotetext{b}{Spectral type and measurement error---we assume that all stars are Dwarf luminosity class V.}
\end{deluxetable}

\subsection{Disk Classification}\label{sec:colors}

We aim to quantify the relative fractions of different disk
morphologies among the A and B stars in W5. The infrared spectral
energy distributions of young stars can show a wide variety of
morphologies, depending on the properties and evolutionary stage of
the disk. In order to display this information, for each star we
calculate the excess as a function of wavelength by deriving:
$(J-[\lambda ])_{dered} - (J - [\lambda ])_{phot}$, i.e. the
dereddened color at wavelength $\lambda$ minus the photospheric
contribution at that wavelength. We assume that there is little or no
excess emission at $J$ band. The photospheric colors for a given
spectral type are calculated by convolving a simple Planck function at
the stellar effective temperature given in \citet{schm82} with the
appropriate filter-plus-telescope-plus-instrument response curves (as
archived for
2MASS\footnote{http://www.ipac.caltech.edu/2mass/releases/allsky/doc/}
and \spitzer\ \footnote{http://ssc.spitzer.caltech.edu/irac/calib/}
\footnote{http://ssc.spitzer.caltech.edu/mips/calib/}). The colors
derived by our blackbody calculation agree with those in the
literature, for example: \citet{bess88}, within 0.05~mag. We use the
infrared extinction law $A_{[\lambda ]}/A_V$ given by \citet{indebe05}
at near-infrared wavelengths and \citet{flaherty07} in the \spitzer\
bands to deredden our infrared photometry for this calculation. We
require that objects possess a 3$\sigma$ excess in at least one filter
to qualify as an excess object. Upper limit detections were not
acceptable as evidence of an excess and we also reject objects that
are visibly non-point-like in either the 8 or 24~$\micron$ images. 

We identify four morphological classes based on the wavelength
dependence of the observed excess emission above theoretical
photospheric levels: (a) optically thick disks that resemble the known
class of Herbig AeBe stars; (b) disks with an optically thin excess
over the wavelength range 2 to 24~$\micron$, similar to that shown by
Classical Be stars; (c) disks that are optically thin in their inner
regions based on their infrared excess at 2--8~$\micron$ and optically
thick in their outer regions based on the magnitude of the observed
excess emission at 24~$\micron$; (d) disks that exhibit empty inner
regions (no excess emission at $\lambda \leq$8~$\micron$) and some
measurable excess emission at 24~$\micron$. A sub-class of disks
exhibit no significant excess emission at $\lambda \leq$5.8~$\micron$,
have excess emission only in the Spitzer 8~$\micron$ band and no
detection at 24~$\micron$.

There already exists a plethora of classifications for disks around
young stars, in particular for low mass stars (M$_\star
\leq$1~M$_\odot$) and the classes we have found in the A and B stars
in W5 should be put into this context. In the low mass case, SEDs are
classified according to their slope in the near to mid infrared into
Class I (an optically thick disk and an envelope), Class II (an
optically thick disk), Class III (no infrared excess) and
`transitional disks' (no infrared excess shortward of
24~$\micron$). \citet{lada06} found a further SED morphology of
optically thin excesses in the IC~348 cluster that they labeled
`anemic' disks. A and B star disks in type (a) resemble the low mass
Class I and II disks. Our type (b) disks are similar to the anemic
disks seen in IC~348. The type (d) disks resemble the `transitional'
disks in low mass stars. The thin/thick disks in type (c) are not
exactly like any of these definitions of disks, but a similar type was
noted by \citet{malfait98} in their survey of nearby Herbig AeBe
stars.

\begin{figure}  
\centering
\begin{tabular}{cc}
\epsfig{file=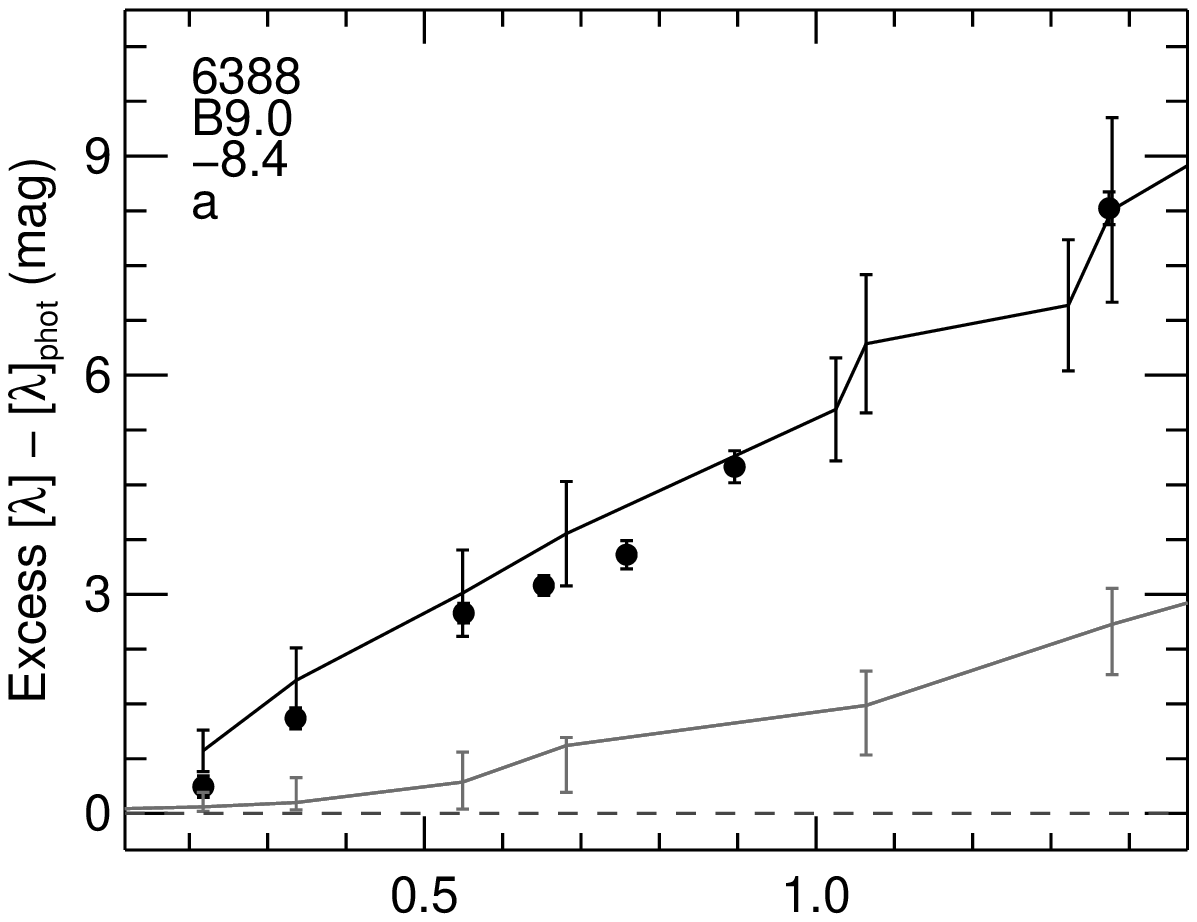,width=0.47\linewidth,clip=} & 
\epsfig{file=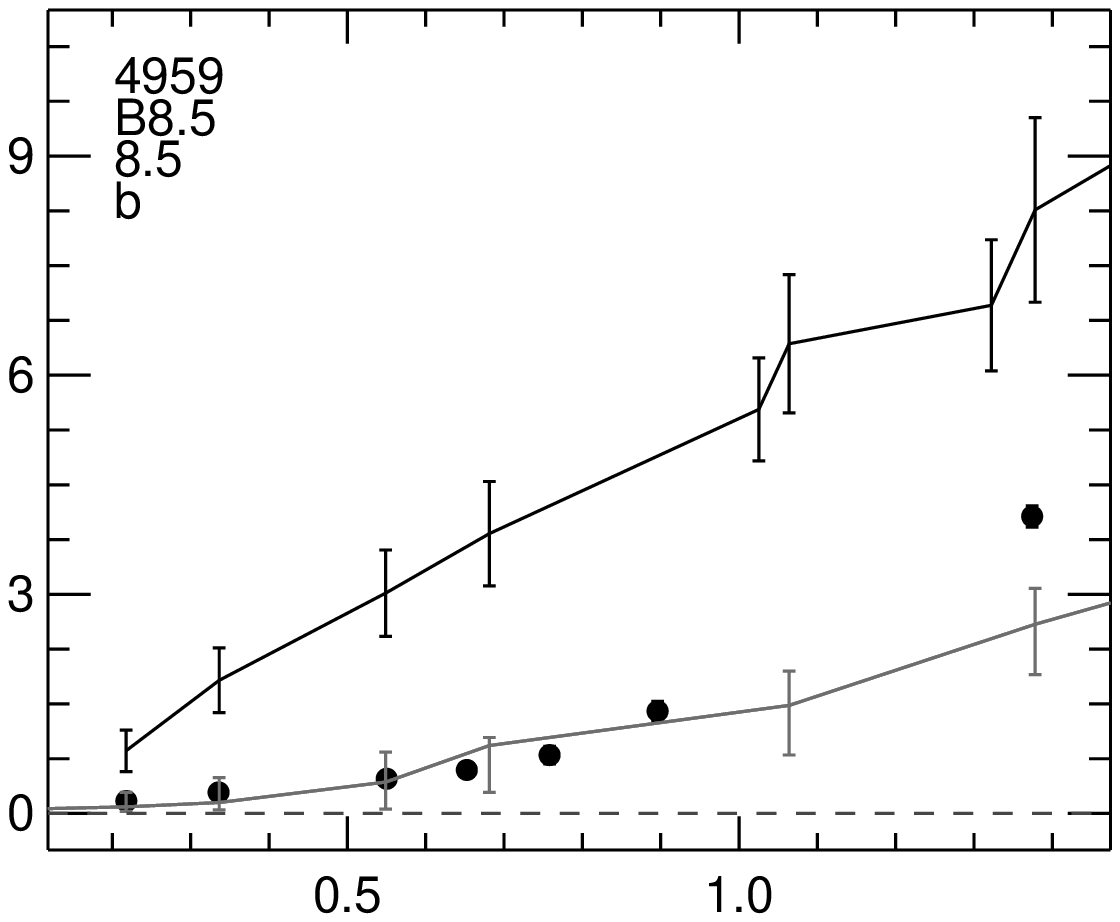,width=0.47\linewidth,clip=} \\
\epsfig{file=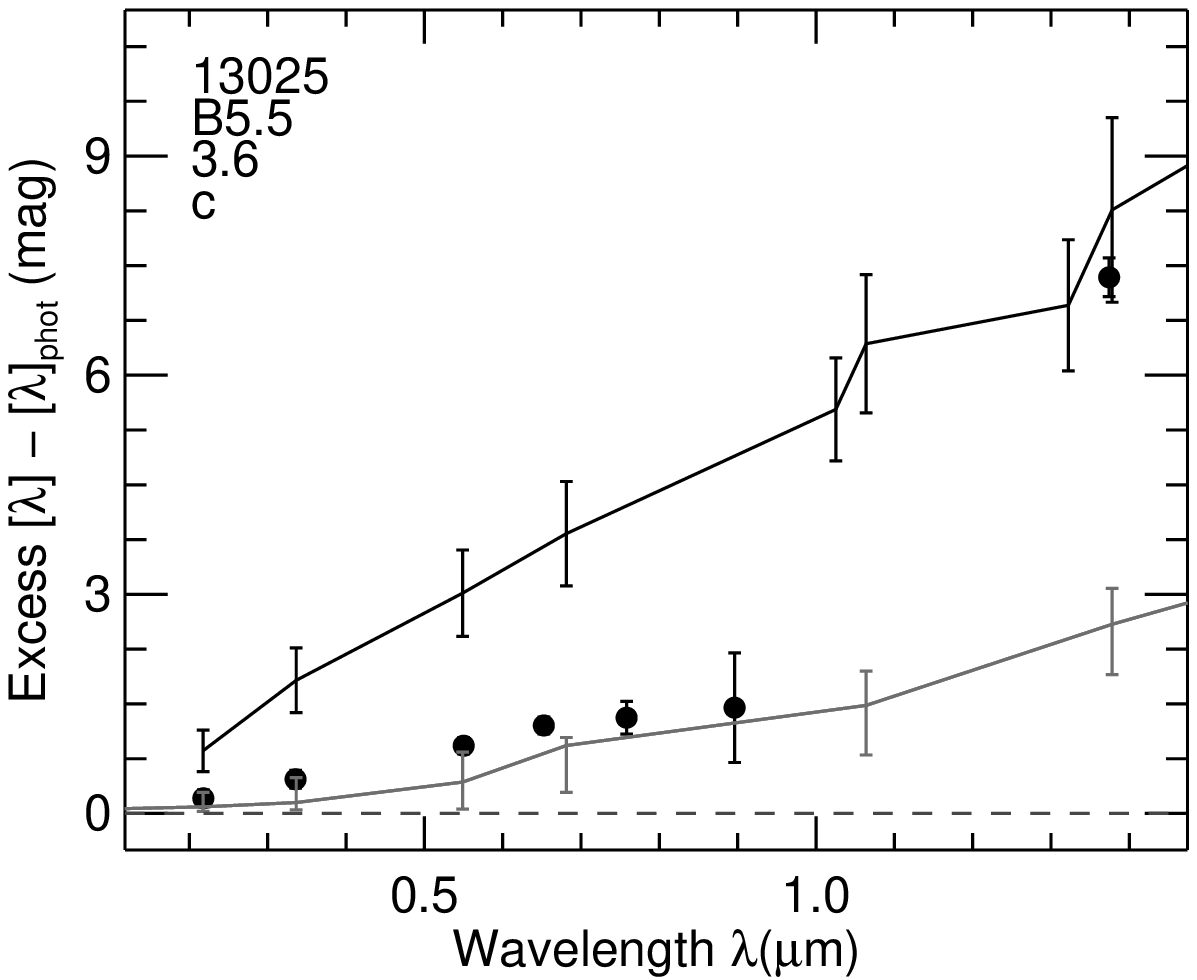,width=0.47\linewidth,clip=} &
\epsfig{file=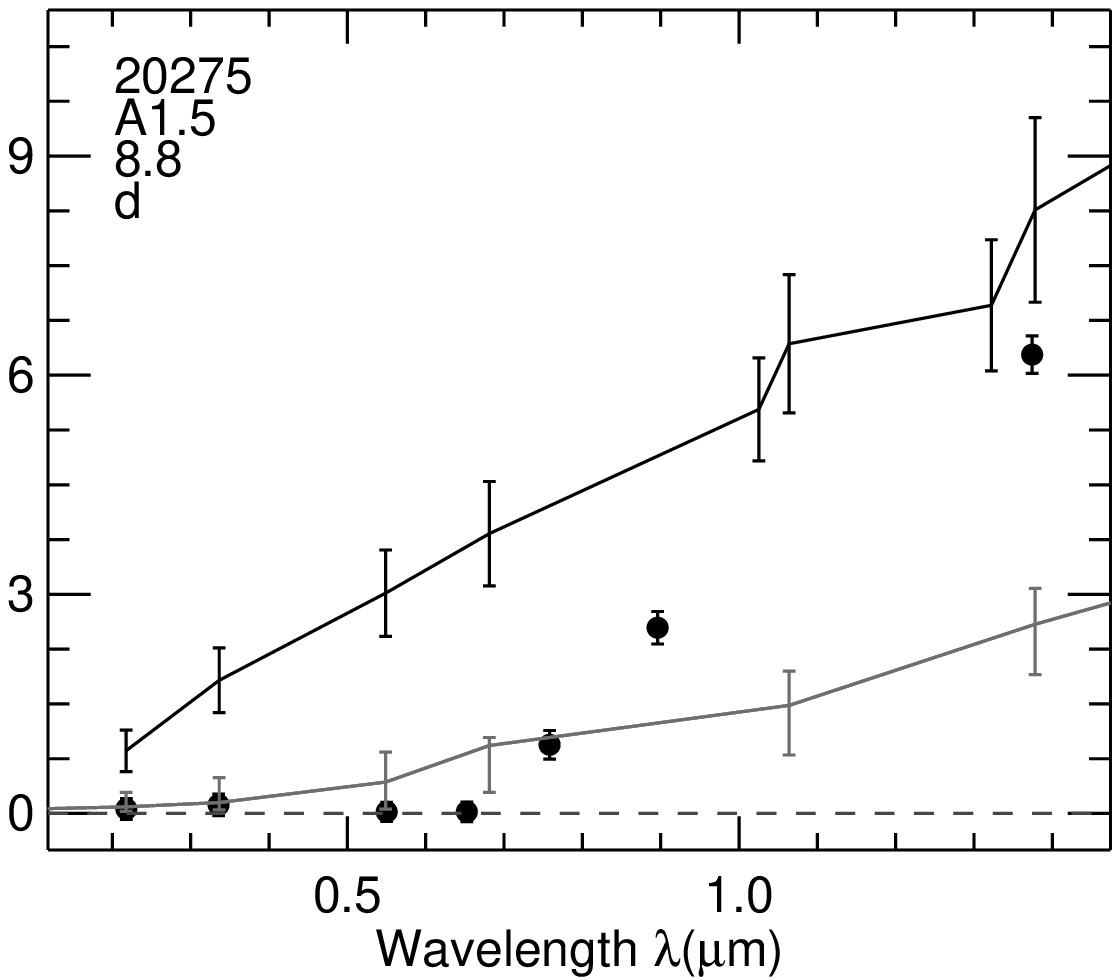,width=0.47\linewidth,clip=}
\end{tabular}
\caption{Distribution of excess emission in magnitudes: $[\lambda
  ]_{dered} - [\lambda ]_{phot}$ as a function of wavelength for four
  example stars (remaining objects are shown in the {\bf
    Appendix}).\label{fig:seds} Error bars include uncertainty in
  $A_V$ calculation. The zero level represents the photospheric level
  for that spectral type. In upper left are shown the spectral type
  and $W({\rm H\alpha})$ in {\rm{\AA}}. Black line: Herbig AeBe star
  median and quartiles from \citet{hillen92}, gray line: CBe star
  median and quartiles from \citet{cote87}, \citet{waters87} and
  \citet{dough91,dough94}.}
\end{figure}

In Figure~\ref{fig:seds} we show the excess emission (in magnitudes)
over the photospheric level as a function of wavelength for four A and
B stars in our spectral sample showing excesses typical of our new
classes. In the list of 610 AB stars in the extinction limited sample
we found 46 (7.5\%) with infrared emission consistent with one of
these four morphologies: Optically thick/Herbig AeBe (a), Optically
thin (b), Optically thin inner and thick outer (c) and `Inner Hole'
(d). The 9 optically thick disks constitute a fraction of 1.5\% of the
extinction limited sample. The remaining stars not grouped under the
four categories above appeared photospheric over the viewable
wavelength range although many objects lacked a detection at 8 or
24~$\micron$, thus we are likely missing some disks. Inner hole disks
are detected around stars over the spectral type range A8.5 to B0 and
mass range 1.6 $<M_{\star} \leq $ 15~$M_{\odot}$.

We present the properties of the 46 spectrally confirmed A and B stars
with IR excess in Table~\ref{tab-exc-only}. IR excess plots for all of
these sources are shown in the Appendix, with the letter corresponding
to the excess class in upper left of each plot. Disks in the sub-class
with excess only at 8~$\micron$ and no 24~$\micron$ detection are
marked as 'de' in the plots.

\begin{deluxetable}{lcccccccccccccccc}
\tabletypesize{\tiny}
\rotate
\tablewidth{0pt}
\tablecaption{AB Stars With IR Excess}
\tablehead{\colhead{ } & \colhead{RA} & \colhead{Dec} & \colhead{$J$} & \colhead{$H$} & \colhead{$K_S$} & \colhead{$[3.6]$} & \colhead{$[4.5]$} & \colhead{$[5.8]$} & \colhead{$[8.0]$} & \colhead{$[24]$} & \colhead{$r'$} & \colhead{$z'$} & \colhead{Sp. Type} & \colhead{EW(H$\alpha$)} & \colhead{$A_V$} & \colhead{ } \\
\colhead{ID\tablenotemark{a}} & \colhead{(deg)} & \colhead{(deg)} & \colhead{(mag)} & \colhead{(mag)} & \colhead{(mag)} & \colhead{(mag)} & \colhead{(mag)} & \colhead{(mag)} & \colhead{(mag)} & \colhead{(mag)} & \colhead{(mag)} & \colhead{(mag)} & \colhead{Type\tablenotemark{b}} & \colhead{($\textrm{\AA}$)} & \colhead{(mag)} & \colhead{Disk\tablenotemark{c}}}
\startdata
9046 & 43.495831 & 60.666310 & 11.00(02) & 10.43(03) & 9.68(02) & 8.49(00) & 8.00(00) & 7.49(00) & 6.93(00) & 5.48(07) & 12.62(12) & 12.24(12) & B8.5$\pm$0.5 & 6.42 & 2.08(50) & a \\
16953 & 45.339983 & 60.482399 & 11.79(03) & 10.80(03) & 9.87(02) & 8.57(00) & 8.08(00) & 7.44(00) & 6.30(00) & 3.69(07) & 14.04(12) & 13.44(12) & B8.5$\pm$0.5 & -17.82 & 2.67(50) & a \\
1562 & 41.952386 & 60.964075 & 12.04(02) & 10.94(02) & 9.96(02) & 8.85(00) & 8.55(00) & 8.08(00) & 7.11(00) & 3.59(07) & 14.06(12) & 13.50(12) & B9.0$\pm$0.5 & -46.50 & 2.48(50) & a \\
5084 & 42.808983 & 60.374804 & 11.68(03) & 11.12(03) & 10.44(02) & 9.12(00) & 8.60(00) & 8.19(00) & 7.70(00) & 6.27(05) & 13.23(12) & 12.90(12) & B9.0$\pm$0.5 & -0.42 & 1.91(50) & a \\
6388 & 43.009641 & 60.604807 & 14.03(04) & 13.35(05) & \nodata & 10.60(00) & 10.16(00) & 9.73(01) & 8.51(01) & 4.96(05) & 16.25(12) & 15.49(12) & B9.0$\pm$0.5 & -8.40 & 3.03(50) & a \\
15405 & 44.933304 & 60.541502 & 13.23(02) & 12.48(03) & 11.60(02) & 10.29(00) & 9.70(00) & 8.97(00) & 7.63(00) & 4.90(04) & 15.11(12) & 14.73(12) & A2.5$\pm$0.5 & -14.19 & 1.80(50) & a \\
5232 & 42.827700 & 60.346598 & 12.50(03) & 12.05(04) & 11.60(02) & 10.65(00) & 10.16(00) & 9.77(01) & 9.09(00) & 7.71(06) & 14.15(12) & 13.73(12) & A6.0$\pm$0.5 & 5.83 & 1.55(50) & a \\
2736 & 42.341200 & 60.751156 & 13.16(03) & 12.11(04) & 11.08(02) & 9.95(00) & 9.55(00) & 9.05(01) & 8.11(02) & 4.23(08) & 15.97(12) & 14.75(12) & A6.5$\pm$0.5 & -11.87 & 3.51(50) & a \\
7717 & 43.254185 & 60.595434 & 12.66(02) & 11.99(03) & 11.34(02) & 10.24(00) & 9.81(00) & 9.20(00) & 7.74(00) & 4.71(08) & 14.55(12) & 14.11(12) & A6.5$\pm$0.5 & -2.40 & 1.52(50) & a \\
13385 & 44.461729 & 60.330359 & 10.71(04) & 10.46(05) & 10.37(04) & 9.89(00) & 9.71(00) & 9.49(01) & 9.25(02) & 7.25(18) & 11.90(12) & 11.66(12) & B0.0$\pm$0.5 & -12.20 & 2.23(50) & b
\enddata
\tablecomments{This Table is published in its entirety in the electronic edition of the {\it Astrophysical Journal}.\label{tab-exc-only} A portion is shown here for guidance regarding its form and content. Values in parentheses by photometry signify error in last two digits of magnitude value. Right Ascension and Declination coordinates are J2000.0. Please contact author for full table.}
\tablenotetext{a}{ID numbering is the same as that from Koenig et~al. (2008). ID numbers greater than 20000 are new to this paper.}
\tablenotetext{b}{Spectral type and measurement error---we assume that all stars are Dwarf luminosity class V.}
\tablenotetext{c}{IR excess morphological type as described in the text. a: Optically thick; b: optically thin; c: optically thin at short $\lambda$, optically thick at long $\lambda$; d: no excess at short $\lambda$, some excess at long $\lambda$.}
\end{deluxetable}

The breakdown of infrared excess types in our spectral sample is
presented in Table~\ref{tab:dtype}. Following the findings of Wolff et
al. (2010, in press) we also note a further sub-category of group (d)
that is: objects with no discernible excess in wavebands up to
5.8~$\micron$ and a real 3$\sigma$ excess at the 8~$\micron$ band. We
find three such objects in the W5 sample and mark them as `de' in
Table~\ref{tab-exc-only}. These may be examples of a disk with a
central hole, or objects with strong PAH line emission (which would
give rise to a sharp increase at 8~$\micron$ but not at shorter
wavelengths).

\begin{figure}
\begin{center}
\epsfig{file=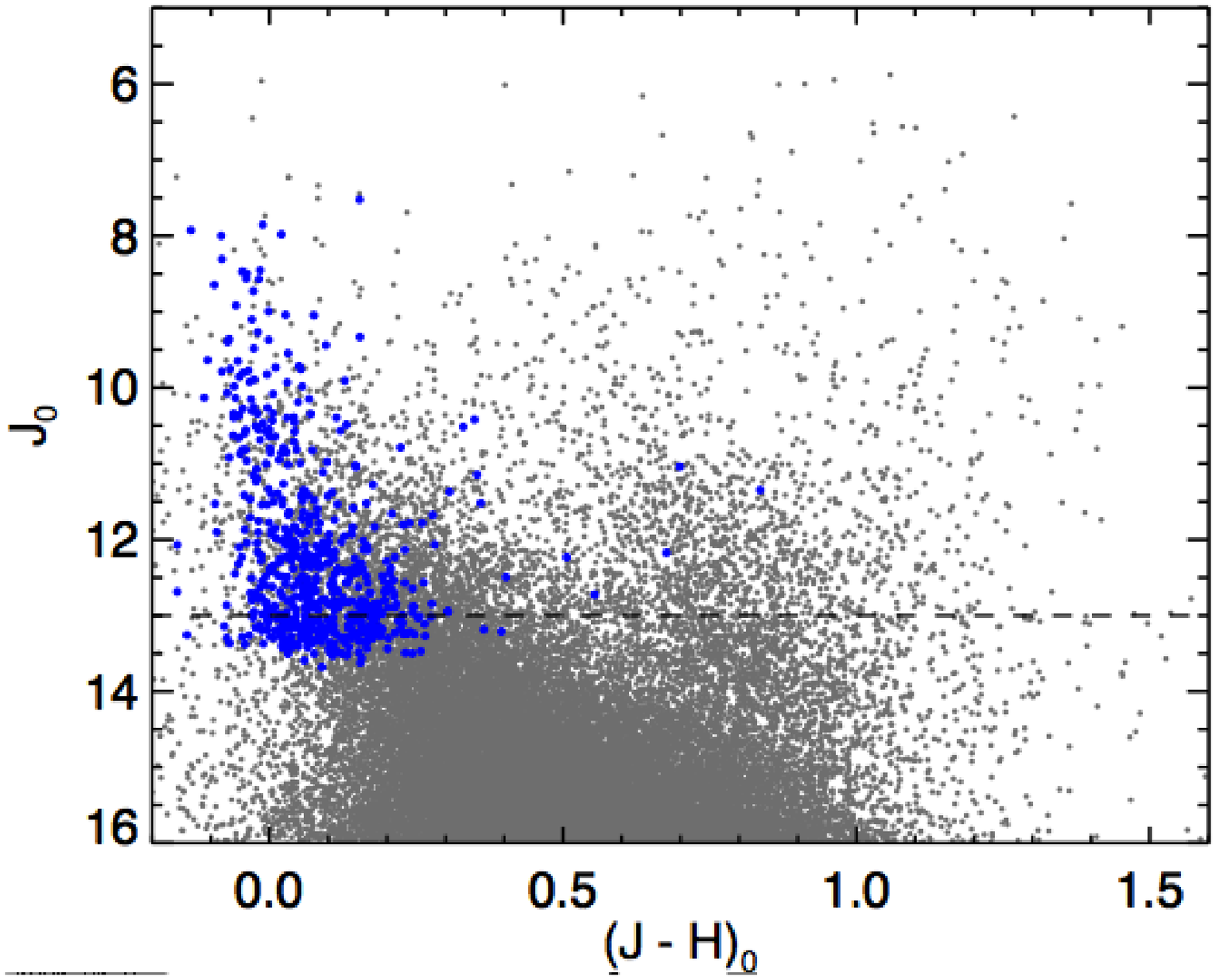,width=0.6\linewidth,clip=}
\caption{Dereddened 2MASS $J$ and $H$ color-magnitude diagram for
  spectroscopically confirmed A and B stars (blue points) and the
  photometric sample in W5 (dark gray points).\label{jjh-phot}
  Photometry is dereddened for spectroscopically confirmed A and B
  stars as in $\S$~\ref{sec-spclass}, and for the photometric points
  with the extinction map of Koenig et~al. (2008). Dashed line shows
  the cut we make in the photometric data to isolate candidate A and B
  stars at the distance of W5.}
\end{center}
\end{figure}

We next expand our sample of intermediate mass stars in W5 by adding
candidate A and B stars from the full $J$ through 24~$\micron$
photometric catalog of \citet{koenig08} that were not included in the
spectroscopic survey. We choose objects with $J$-band and IRAC channel
1 to 4 photometric uncertainty $\leq$0.2, a total of 14283 stars,
excluding the list for which we have already obtained spectra. As
described by Koenig et~al. we estimate the extinction for each source
in the absence of a spectral type from an $A_V$ map constructed with
2MASS $HK_S$ photometry of sources across W5. We deredden the
photometry with the value of $A_V$ given by the location of each
source within the map. We assume that A and B stars at the distance of
W5 will have a dereddened $J$ magnitude $\leq$13 and use this limit as
a first cut to find them in the photometric sample: a total of 4805
candidate A and B stars. In Figure~\ref{jjh-phot} we show a dereddened
plot of 2MASS $J-H$ versus $J$ for the spectroscopic sample of A and B
stars (with extinction as calculated in $\S$~\ref{sec-spclass}) and
for the photometric sample in W5 dereddened with the extinction map as
described in \citet{koenig08}.

\begin{figure}
\centering
\begin{tabular}{cc}
\epsfig{file=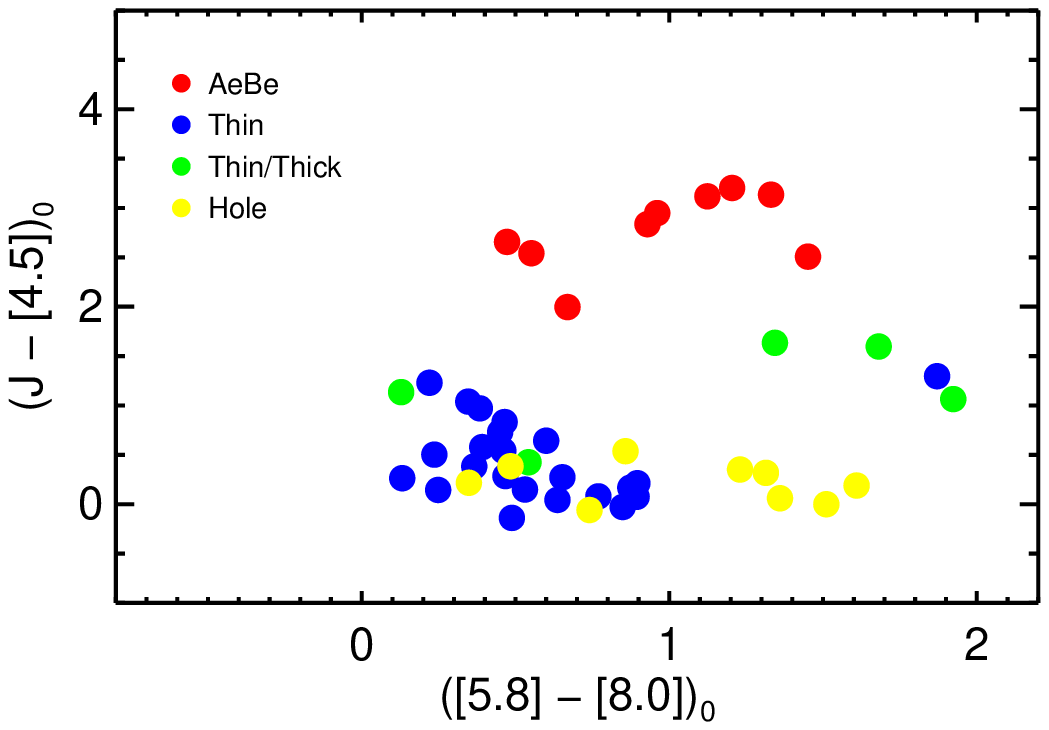,width=0.5\linewidth,clip=} &
\epsfig{file=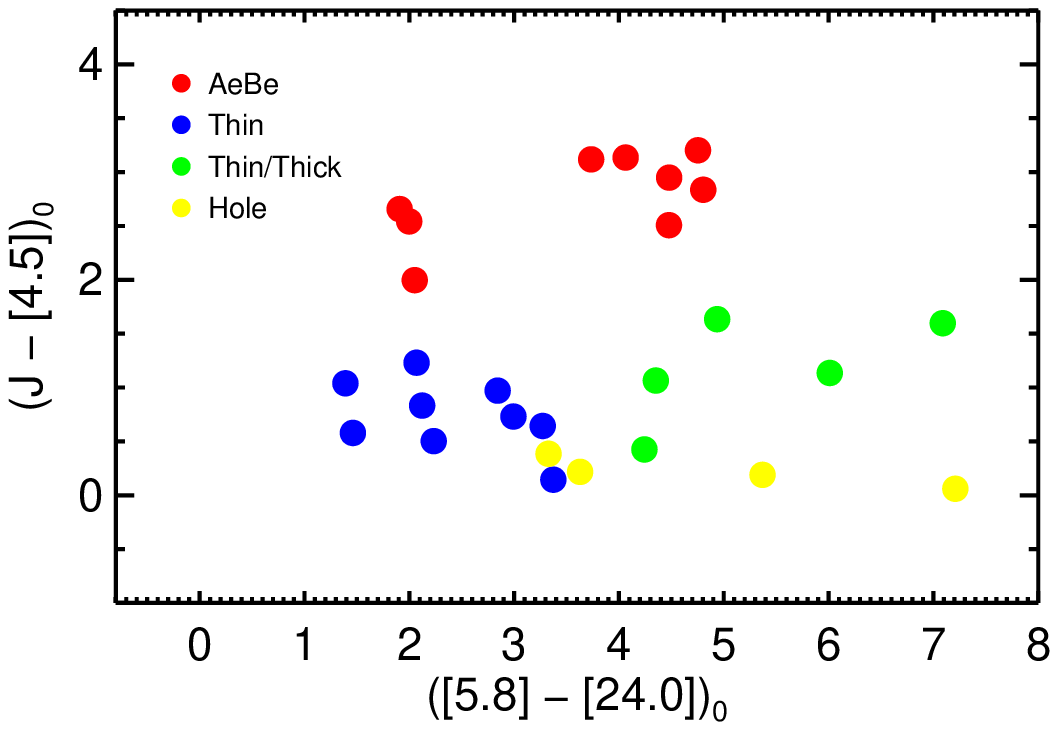,width=0.5\linewidth,clip=} \\
\epsfig{file=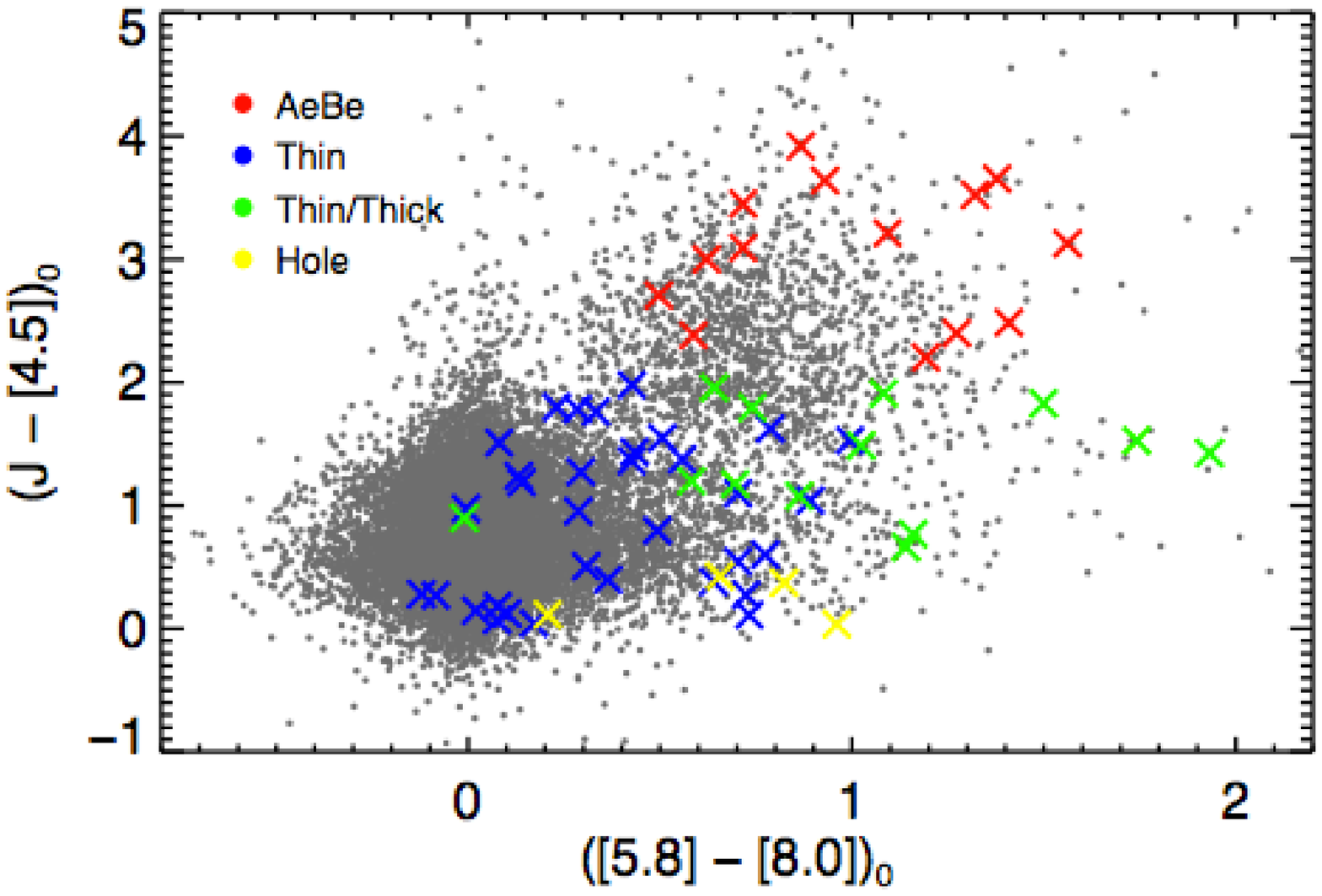,width=0.5\linewidth,clip=} & 
\epsfig{file=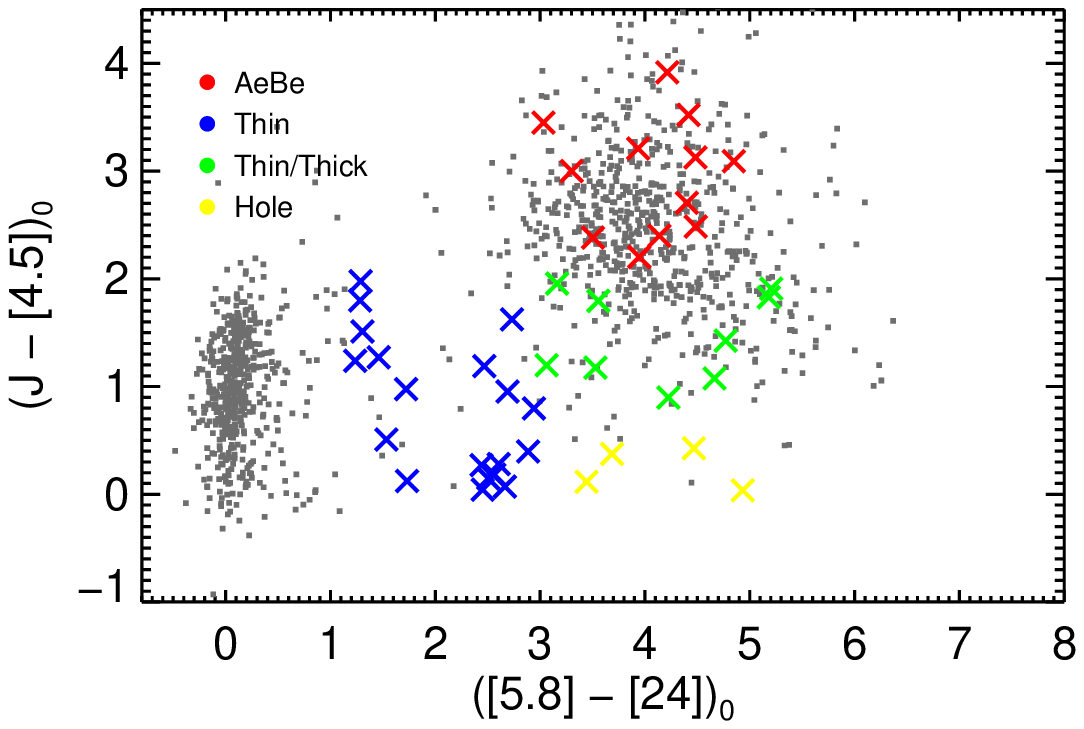,width=0.5\linewidth,clip=} 
\end{tabular}
\caption{Upper panels: Color-color diagrams for the spectroscopic
  sample of A and B stars in W5 possessing at least a 3 sigma excess
  in one or more bands, color-coded according to disk
  class.\label{fig:ccds} Red: type (a), blue: type (b), green: type
  (c), yellow: type (d). Lower panels: Color-color diagrams for the
  photometric sample in W5: gray points show all photometric
  detections, colored crosses show best candidate A and B stars with
  the same color scheme as for the spectroscopic sample.}
\end{figure}

The $J_0$=13 limit roughly corresponds to a main-sequence star of
spectral type A5 at 2~kpc \citep{schm82,kenyon95}. With the spectral
sample as a guide, we examine the infrared colors of sources brighter
than this and classify them into the same IR SED groups by their
location in color space in Figure~\ref{fig:ccds}. We only include
objects with an excess in at least one color: a total of 64 objects,
as before excluding stars for which we already have spectra. The full
classification scheme is described in the
Appendix. Table~\ref{tab:dtype} presents the resulting breakdown of
disks in the photometric sample by type. The candidate photometric A
and B stars with IR excess are plotted in Fig.~\ref{fig:ccds} (lower
panels).

The method of finding the photometric-only sample of A and B stars is
crude since we only require they be bright and possess an apparent
infrared excess. As a test, we applied the same criteria to our
spectroscopic catalog of 4800 objects and found 60 objects. Of these
stars 62\% proved to be A or B type, since some later type stars can
be similarly bright and possess excess infrared emission. The
contamination fraction is not constant between the different disk
types. The fraction of confirmed A and B stars by group was: Thick:
50\%, Thin: 63\%, Thin/Thick: 45\%, Hole: 83\%. We add a further row
to Table~\ref{tab:dtype} applying these success rates to the
photometric sample. We note that the proportions of the different disk
types are consistent within the errors.

\begin{deluxetable}{lcccc}
\tablewidth{0pt} 
\tablecaption{Disk Type Breakdown}
\tablehead{ \colhead{ } & \colhead{HAeBe} & \colhead{Thin} & \colhead{Thin/Thick} & \colhead{Hole} }
\startdata
Spectral & 9 & 23 & 5 & 9 \\
Photometric & 14 & 33 & 13 & 4 \\
\hline {\bf Total} & 23 & 56 & 18 & 13 \\
(\%) & 21$\pm$4.8 & 51$\pm$8.4 & 16$\pm$4.2 & 12$\pm$3.5 \\
{\bf Alt. Total}\tablenotemark{a} & 16 & 44 & 11 & 12 \\
(\%) & 19$\pm$5.3 & 53$\pm$9.9 & 13$\pm$4.3 & 15$\pm$4.5\label{tab:dtype}
\enddata
\tablenotetext{a}{Alternative totals assuming contamination of
  photometric sample by late type stars of 50\%, 37\%, 55\%, 17\%
  respectively. Errors in percentages are derived assuming Poisson
  statistics.}
\end{deluxetable}

\subsection{Accretion Signatures in the AB star sample}
The spectra in this paper have insufficient resolution to precisely
measure H$\alpha$ equivalent widths (EW(H$\alpha$) and are subject to
large errors introduced by our crude sky subtraction technique. Thus
we are unable to establish reliable measures of the accretion rates in
our sample of A and B star disks. However, we can still use the
measurements of EW(H$\alpha$) from our spectra to assess to first
order which of the stars exhibit some noticeable excess emission and
are still accreting.

\begin{figure}
\begin{center}
\epsfig{file=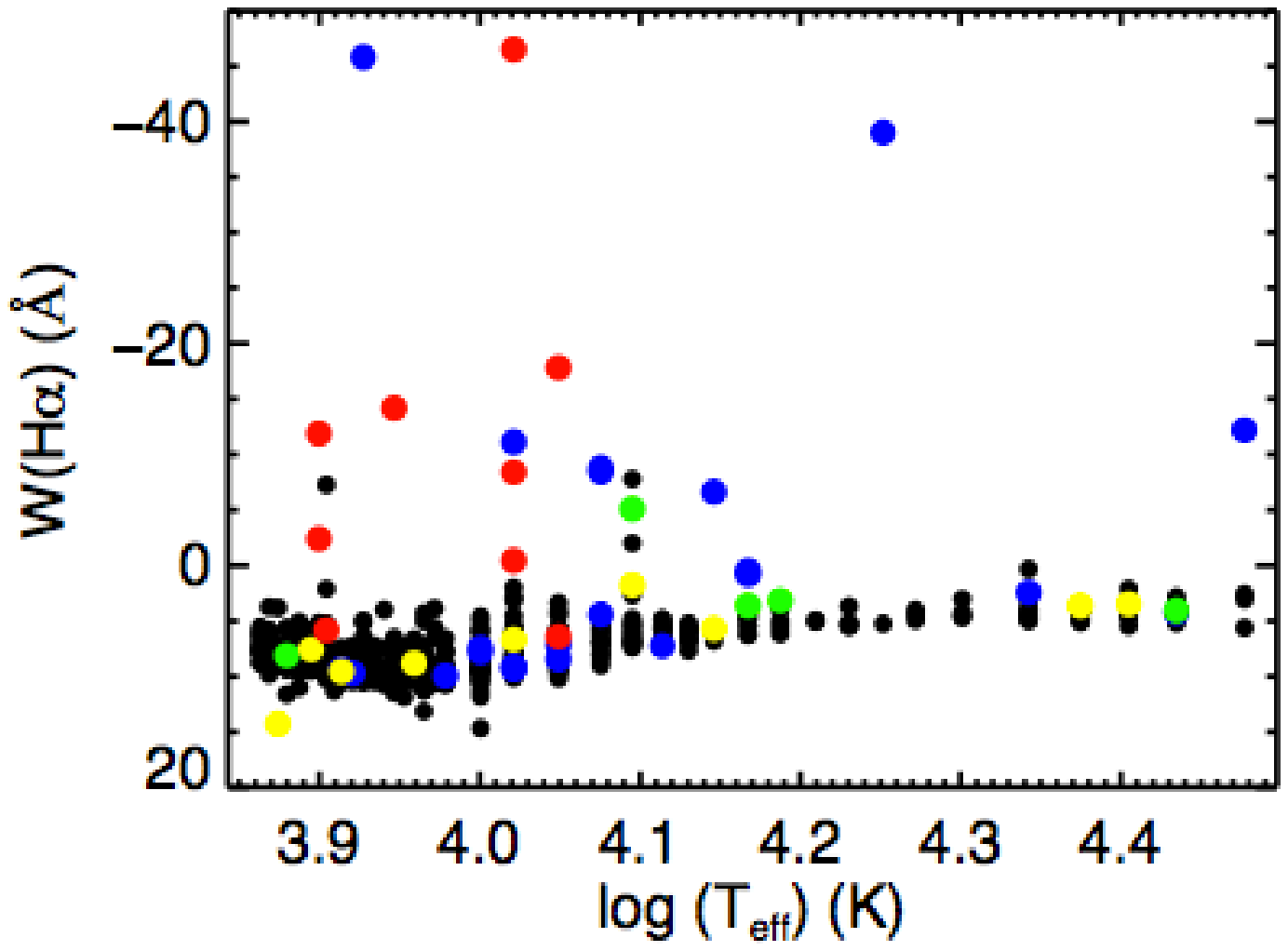,width=0.6\linewidth,clip=}
\caption{H$\alpha$ equivalent width as a function of effective
  temperature for the A and B stars in our sample.\label{fig:halpha}
  Objects showing disk excess are color-coded according to the color
  scheme established in Figure 6.}
\end{center}
\end{figure}

In Figure \ref{fig:halpha} we plot the measured EW(H$\alpha$) as a
function of effective temperature for all of the A and B stars in the
spectral sample. We overlay on this plot the disk excess objects,
color-coded according to our earlier color scheme. Negative values of
EW(H$\alpha$) indicate emission lines.

The Main Sequence level of H$\alpha$ absorption is clearly traced from
$\sim$10~{\AA} in the A stars below 10000~K to $\sim$5~{\AA} in the
hottest B stars at 30000~K. The largest {\it negative} values of
EW(H$\alpha$) (corresponding to emission) are in the thick disk and
thin disk sources. In the case of the thick disk sources this emission
is likely produced by accretion from the circumstellar disk. One of
the `thick' objects (ID 5267, a B5.5 star) was observed at multiple
epochs (7-Oct-2007 and 30-Sept-2008). Its H$\alpha$ equivalent width
varied from -7.99 to 0.49~{\AA} between these two observations,
suggesting variable accretion activity in this source. The `thin'
sources' SEDs resemble those of Classical Be stars which exhibit
strong emission in this line produced in a stellar wind
\citep{porter03}. The thin/thick and hole sources also exhibit some
evidence for H$\alpha$ in emission which suggests that some of these
stars may still be accreting material from their disks but clearly at
a consistently lower rate than in the thick disks.

\subsection{Spatial Variation in SED types}
\begin{figure}
\begin{center}
\includegraphics[angle=90,width=5.7in]{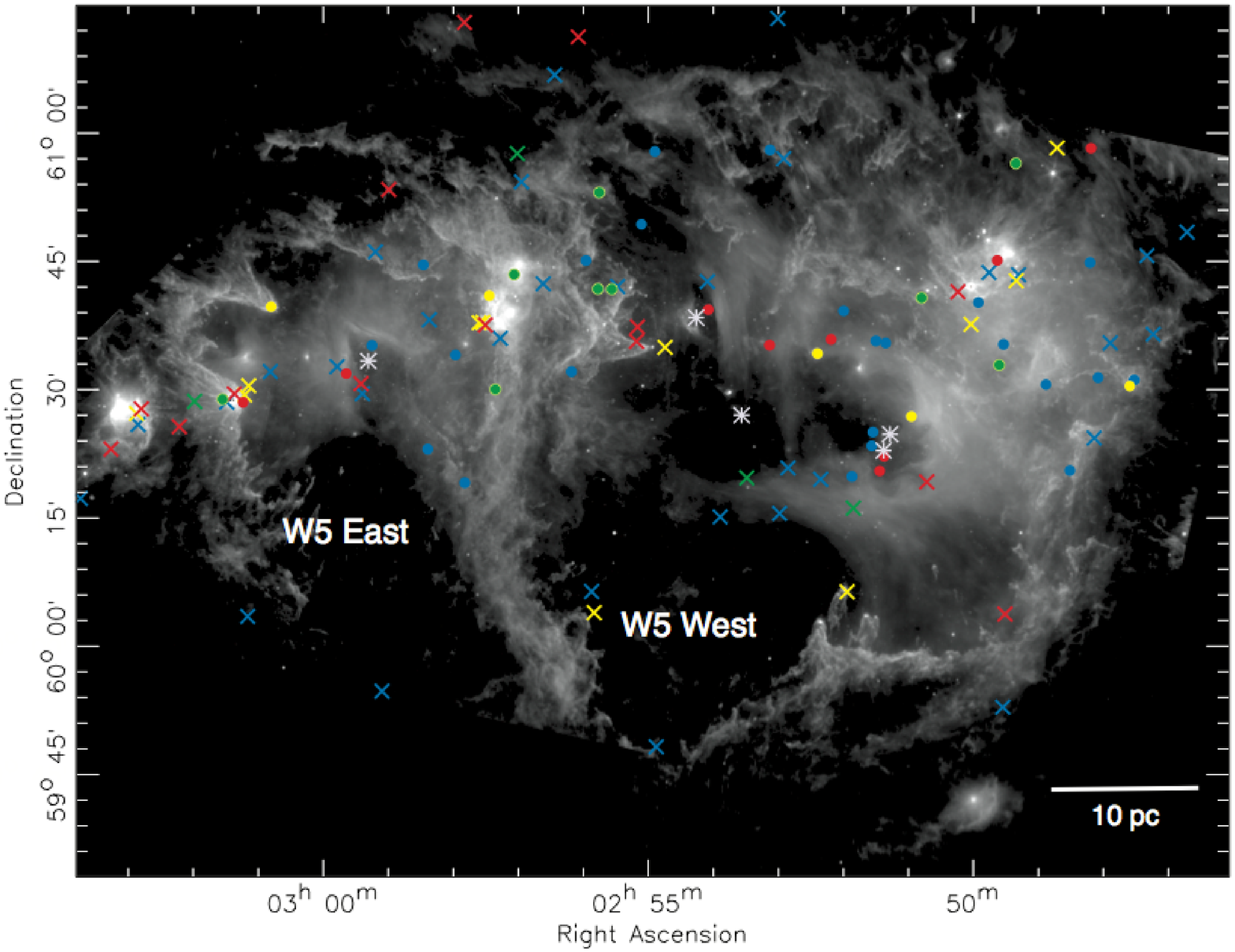}
\caption{{\it Spitzer} MIPS 24 $\micron$ mosaic overlaid with AB star
  disk candidates. Dot points show spectroscopically confirmed A and B
  stars.\label{fig:distrib} Cross points mark photometric-only
  sources. Color-coding is as in Fig.~\ref{fig:ccds}. Known O stars in
  W5 are marked with white asterisks.}
\end{center}
\end{figure}

We present the spatial distribution of A and B star disks in W5 in
Figure~\ref{fig:distrib} on the \spitzer\ MIPS 24~$\micron$ mosaic. We
find no systematic trends of location with disk type, however we note
that the high extinction on the molecular clouds (marked by the bright
diffuse 24~$\micron$ emission around the rim of W5) will strongly
reduce the number of disk excess objects we find in our survey in
these regions. In \citet{koenig08} we calculated the 90\% completeness
limits for our \spitzer\ survey as a function of location and
wavelength. In the \htwo\ region cavities we find 90\% limits of 16.1
(4.5~$\micron$), 14.6 (5.8~$\micron$), 13.4 (8~$\micron$) and 11.2
(MIPS, 24~$\micron$). In the bright diffuse emission regions, the
limits are: 14.2 (4.5~$\micron$), 12.3 (5.8~$\micron$), 9.5
(8~$\micron$) and 8.5 (MIPS, 24~$\micron$). Adopting theoretical
colors as calculated in $\S$\ref{sec:colors} these results mean we are
able to detect all A and B star excess types in the \htwo\ region
cavities. On the cloud however, we are limited by a convolution of the
detection limits in each band with the range of excesses for each disk
type. Table~\ref{tab:comp} presents a summary of the resulting
completeness limits. As a consequence we miss almost all A-type Thin
and Hole disk sources against the bright cloud diffuse emission.

\begin{deluxetable}{lcccc}
\tablewidth{0pt}
\tablecaption{Disk Type Completeness}
\tablehead{ \colhead{ } & \colhead{Thick Disks} & \colhead{Thin Disks} & \colhead{Thin/Thick Disks} & \colhead{Hole Disks}}
\startdata 
90\% Completeness Limit & A9.5 & A0 & A9.5 & A0\\
Brightest 50\% Only & N/A & A1 & N/A & A4.5\label{tab:comp}
\enddata
\end{deluxetable}

\section{Discussion}\label{sec:discuss}
We have found a sample of young A and B stars in W5 that possess a
variety of disk morphologies and that the disks can be grouped into
four distinct types. Although this scheme is a crude way to describe
what is almost certainly a continuum of disk morphologies, we can use
these types to investigate the properties and origins of these
disks. Indeed a similar scheme was constructed by \citet{malfait98}
from their observations of near and mid-infrared excesses in nearby A
and B stars. We have been able to note this same range of disk types
in a single young region.

We firstly assume that all stars of intermediate mass begin their life
in a protostellar phase similar to that of low mass stars. As such,
they initially possess an accretion disk and an infalling envelope. As
time progresses, the envelope disappears and an optically thick
accretion disk is left behind. \citet{andrews07} and
\citet{muzerolle00} have shown that in low mass stars, the
disk-to-star mass ratio can span two orders of magnitude (from 0.1 to
10\% of the stellar mass) as can the disk accretion rate onto the
star. If the same property applies to intermediate mass stars, then
the initial conditions alone can produce very different end states
after a given amount of time in the AB stars' disks in a cluster as
these disks drain onto their stars and dissipate through spreading.

Viscous draining and spreading of a disk alone are likely too slow to
explain observed disk evolutionary timescales
\citep{hartmann98}. Several additional mechanisms exist that will
alter the evolution of the star-disk system in time and operate on
more rapid timescales. The disk may photoevaporate due to radiation
from the host or external stars. Dust grains in the disk can settle to
the mid-plane and grow in size. Finally, large solid bodies
(planetesimals and planets) may form in the disk. There is evidence
that disk evolution proceeds faster in more massive stars than in low
mass stars \citep{haisch01}. \citet{hern05} suggest that the timescale
for removal of the optically thick inner disk in Herbig Ae/Be stars
(types B5 to A9) is short, given that the fraction of stars with an
optically thick inner disk as seen in the $JHK_S$ bands is lower at a
given age amongst A and B stars than in low mass GKM stars. In W5 we
find no optically thick disks in our spectral sample in stars earlier
than B8.5. Put another way, only 20$\pm$5\% of stars with any kind of
excess are still optically thick in the spectroscopic-only sample, and
none above a mass 2.4~$M_\odot$.

External erosion of low mass star disks has been clearly observed in
the Orion Nebula near the Trapezium stars \citep{odell93}, but the
expected photoevaporation rate produced drops off exponentially with
distance if the star and disk are beyond 1~pc from the powering O
star. Photoevaporation of the disk by its host star is driven by
heating of the disk surface by X-ray (h$\nu >$0.1~keV),
Far-ultraviolet (FUV, h$\nu <$13.6~eV) and Extreme ultraviolet (EUV,
h$\nu >$13.6~eV) radiation. FUV radiation heats the disk. Beyond a
certain radius in the disk, the gas thermal velocity exceeds the
escape velocity of the system and material can leave in a
photoevaporative wind. This effect removes the outer disk while the
inner disk drains onto the star through accretion. X-rays and EUV
radiation may be able to create a gap in the disk and subsequently
rapidly clear the entire inner disk \citep{clarke01}. Key to the
operation of this mechanism is the ability of the EUV and X-ray
photons to penetrate the accretion flow and ionize the disk---a still
open question \citep{ercolano08}. A disk cleared out from the inside
in this manner might resemble the `inner hole' sources in our
survey. The timescale for this process is short however, typically
less than 10$^5$~yr \citep{alexander06}.

As a disk evolves, dust grains will settle to the disk mid-plane and
can grow in size. Grain growth and planetesimal formation will reduce
the optical depth in small grains which could explain the reduced
emission in the inner disk as seen in the thin/thick sources. Gas will
still be present in such a disk and so the accretion rate should
remain at a similar level as in the optically thick disks. The
formation of a giant planet in the disk (M$\approx$M$_{Jup}$) could
also produce the thin/thick morphology. The planet both reduces the
dust optical depth and decreases the accretion rate of material onto
the star by slowing the rate of flow of material from the outer to the
inner disk \citep{rice06}. In this case the accretion rate will be
around 10\% of the typical thick disk rate. A large enough planet
(M$\gg$M$_{Jup}$) can completely clear the inner disk, resulting in a
disk more like the `hole' sources. A study of the disks around low
mass stars in Taurus appears to show that stars with IR morphologies
similar to our thin/thick sources do indeed have accretion rates about
10$\times$ lower than the optically thick accretion disks
\citep{najita07}. If the same is true for the AB stars in W5, then
these sources may be good candidates for giant-planet forming
disks. The H$\alpha$ line in the thin/thick sources are systematically
smaller than in the thick sources in Fig.~\ref{fig:halpha}, however
these results are not precise enough to establish their relative
accretion rates. The hole sources are similar to the thin/thick disks
in that they show some process that has cleared out the inner
disk. These could be examples of giant planet formation as well, but
could also simply be photoevaporated disks. Again their H$\alpha$
lines show less emission than the thick disk sources which could
indicate lower accretion rates in these objects.

\begin{figure}
\begin{center}
\includegraphics{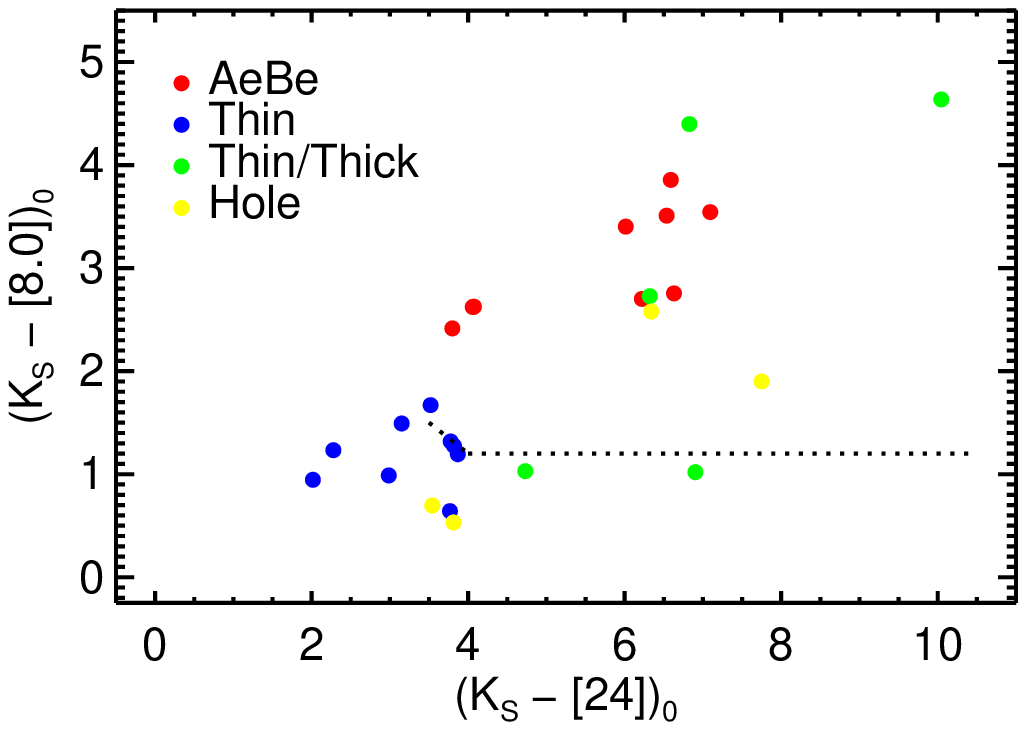}
\caption{$K_S - [8.0] vs. K_S - [24]$ color-color plot of
  spectroscopic A and B stars in W5, analogous to Figures 18 to 25 in
  Luhman et~al. (2010).\label{fig-luhman-comp} The lower boundary of
  primordial disks in Taurus that they identify is shown as a dotted
  line.}
\end{center}
\end{figure}

The thin disk objects in our sample resemble so-called homologously
depleted or weak disks. These disks may represent the action of grain
growth or some process that drives mass loss from the disk at a wide
range of disk radii. \citet{luhman10} and \citet{muzerolle10}
discovered similar disks in their recent studies of low mass young
stars, but in far fewer numbers. In Figure~\ref{fig-luhman-comp} we
make a plot similar to their Figures 18--25. They typically find $<$10
objects below the boundary marking the location of Taurus primordial
disks (with the exception of the sample from IC 348). As a fraction of
all disks, our sample of thin disks in W5 is larger than in any of
their low mass disk samples. We do not know at present why they are
more prevalent among young intermediate mass stars. The earliest types
showing this excess are a B0, a B0.5 and a B2 star.

\section{Conclusions and Future Work}
We have carried out a combined infrared and optical photometric and
spectroscopic survey of young stars in W5. We present evidence from
the larger spectroscopic sample that W5 is a young region (age
$\lesssim5$~Myr), in agreement with the presence of massive O stars
and a short expansion timescale for the \htwo\ region bubbles. More
detailed analysis of the HR diagram would help to place better
constraints on the age of W5.

Within the spectroscopic sample we identify a population of A and B
stars. Of these, we find that 46 objects show excess infrared emission
in at least one \spitzer\ band. We find that many of these infrared
excess sources suggest evolved disks with little or no excess infrared
emission above the photosphere up to a wavelength 5.8--8~$\micron$,
but strong excess at longer wavelengths. These may be transitional
disk objects that have cleared out their inner disks. Photoevaporation
is one possible mechanism to create this SED morphology, however there
is a distinct possibility that their SEDs are produced by the
formation of one or more planets. If we can assess the contribution of
the various mechanisms of disk evolution and their frequency of
occurence, we can begin to understand what types of planets will form
in a given disk around a star of given mass, how frequently planet
formation occurs in the Galaxy and at what point in a star's evolution
it begins. A clearer picture of the planet forming potential of
intermediate mass stars should soon emerge.

\acknowledgements The authors would like to thank Steve Strom and
Sidney Wolff, Joan Najita, James Muzerolle, Lee Hartmann and Nuria
Calvet, Uma Gorti and Dave Hollenbach for extensive and useful
discussions on disks and young stars. This work is based (in part) on
observations made with the {\it Spitzer} Space Telescope, which is
operated by the Jet Propulsion Laboratory, California Institute of
Technology under a contract with NASA. Support for this work was
provided by NASA. This publication makes use of data products from the
Two Micron All Sky Survey, which is a joint project of the University
of Massachusetts and the Infrared Processing and Analysis
Center/California Institute of Technology, funded by the National
Aeronautics and Space Administration and the National Science
Foundation. This research has made use of NASA's Astrophysics Data
System. This research has made use of the SIMBAD database, operated at
CDS, Strasbourg, France.

{\it Facilities:} \facility{MMT(Hectospec, Megacam)}, \facility{FLWO:1.5m(FAST)}, \facility{FLWO:1.2m(Keplercam)}, \facility{Spitzer (IRAC, MIPS)}

\appendix
\section{Disk Classification Scheme}
We classify disks from the photometric sample of stars possessing
dereddened $J$ magnitudes brighter than 13 according to the following
scheme using dereddened photometry in all cases. Firstly we identify
Herbig AeBe star candidates.

\begin{equation}
J-[4.5] > 2.0
\end{equation}
\begin{equation}
[5.8]-[8.0] > 0.3
\end{equation}
\begin{equation}
J-H > \frac{0.7}{0.55}(H-K_S)-0.6
\end{equation}
\begin{equation}
J-H > -\frac{0.25}{0.28}(H-K_S)-0.59
\end{equation}
\begin{equation}
J-H > \frac{0.95}{0.55}(H-K_S)-0.116
\end{equation}
\begin{equation}
J-H < 1.2
\end{equation}

Next we identify the `thin' disk candidates:

\begin{equation}
J-[4.5] > -4.5([5.8]-[8.0])+3.0
\end{equation}
\begin{equation}
-0.25 < J-[4.5] \leq 2.0
\end{equation}

and either:

\begin{equation}
[5.8]-[8.0] \leq 1.0
\end{equation}

or:

\begin{equation}
1.2 < [5.8]-[24.0] \leq 3.0
\end{equation}

We next pick out the `thin/thick' sources. In some cases this
re-classifies thin-disk sources.

\begin{equation}
0.5 < J-[4.5] \leq 2.0
\end{equation}

and either:

\begin{equation}
[5.8]-[8.0] > 1.0
\end{equation}

or:

\begin{equation}
[5.8]-[24.0] > 3.0
\end{equation}

Lastly we find the `hole' sources:

\begin{equation}
-0.25 \leq J-[4.5] \leq 0.5
\end{equation}

and either:

\begin{equation}
[5.8]-[8.0] > 1.0
\end{equation}

or:

\begin{equation}
[5.8]-[24.0] > 3.0
\end{equation}

\section{IR Excess Plots}
\begin{figure}
\begin{center}
\epsfig{file=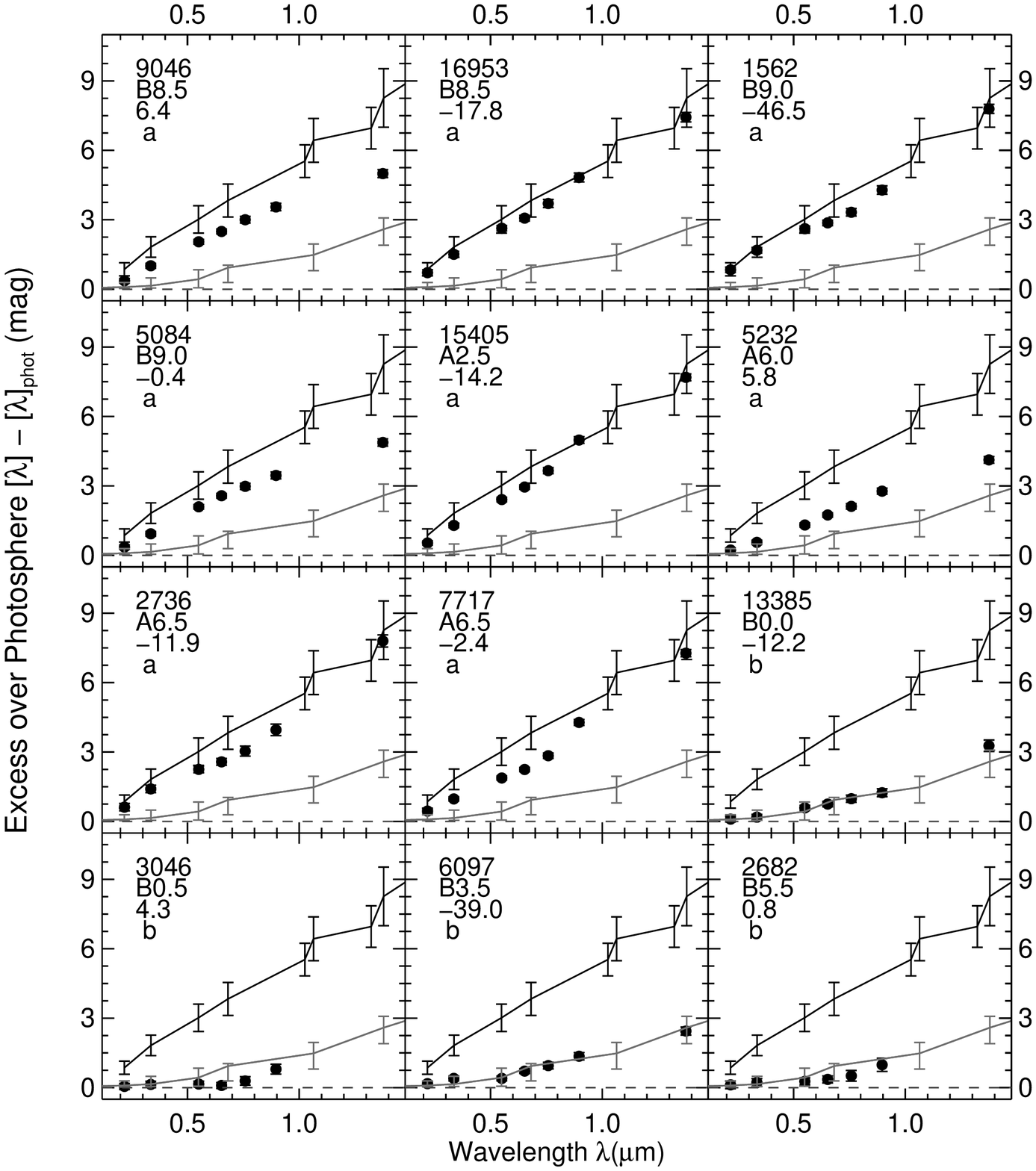,width=0.99\linewidth,clip=}
\caption{IR Excess plots for the spectrally confirmed A and B stars
  with infrared excess.\label{excess-all} Black points show the data
  and 1$\sigma$ error bars. Upper line and error bars shows Group 1
  Herbig AeBe star median and quartiles from \citet{hillen92}, lower
  line and error bars shows the Classical Be star median and quartiles
  from \citet{cote87}, \citet{waters87} and
  \citet{dough91,dough94}. In upper left of each plot is given the
  source ID, spectral type, EW(H$\alpha$) and disk type.}
\end{center}
\end{figure}

In Figure~\ref{excess-all} we display IR excess plots for all of the
remaining 42 spectrally confirmed A and B stars that belong to groups
a, b, c or d. Sources are listed by group, and within each of these
divisions are sorted by spectral type. Every plot also is overlaid
with the Group I Herbig AeBe star median and quartiles from
\citet{hillen92}, and a line showing the Classical Be star median and
quartiles from \citet{cote87}, \citet{waters87} and
\citet{dough91,dough94}, where the excesses used to find the medians
are calculated in the same way as for the W5 stars.

\begin{figure}
\begin{center}
\epsfig{file=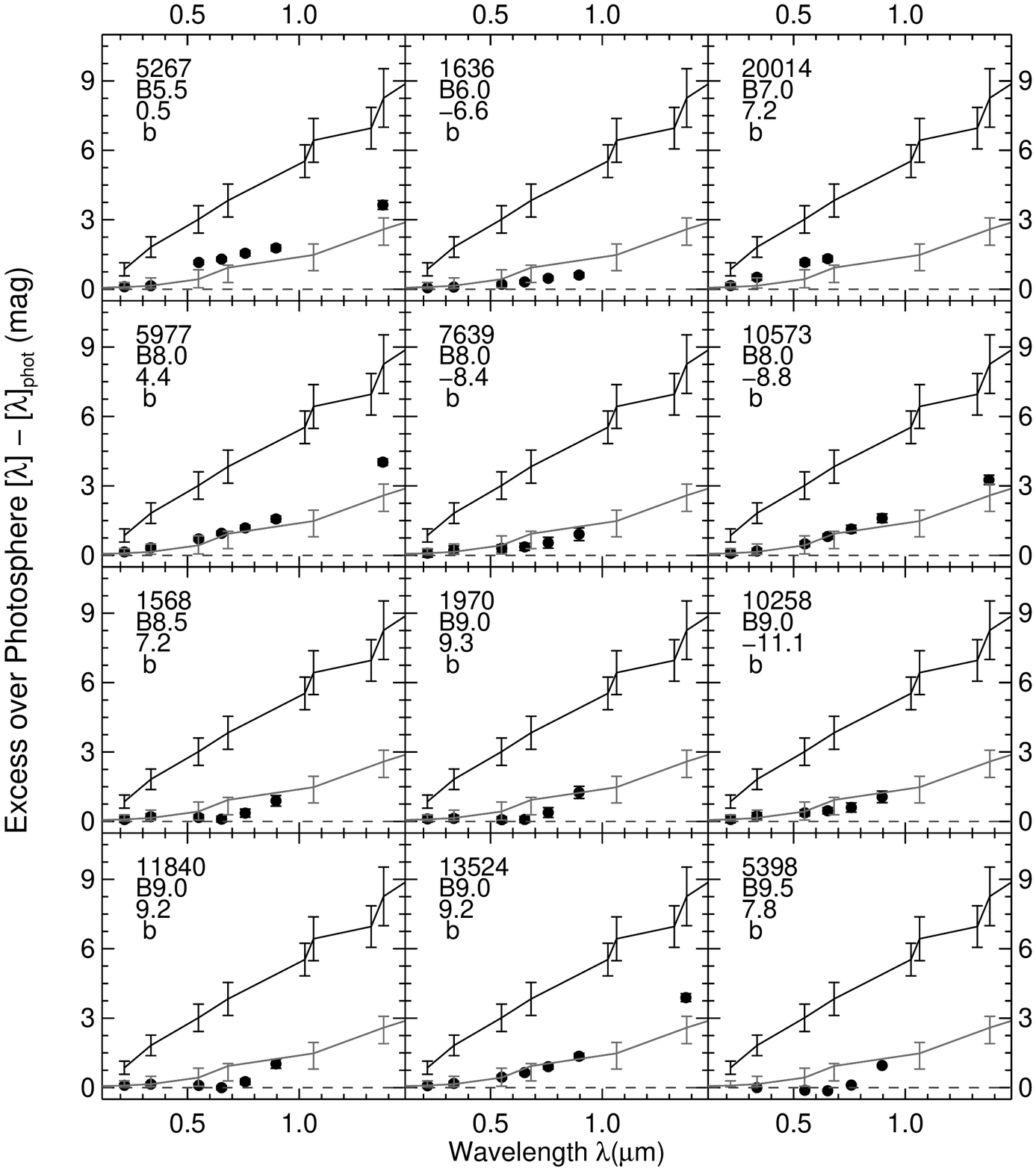,width=0.99\linewidth,clip=}
\end{center}
\end{figure}

\begin{figure}
\begin{center}
\epsfig{file=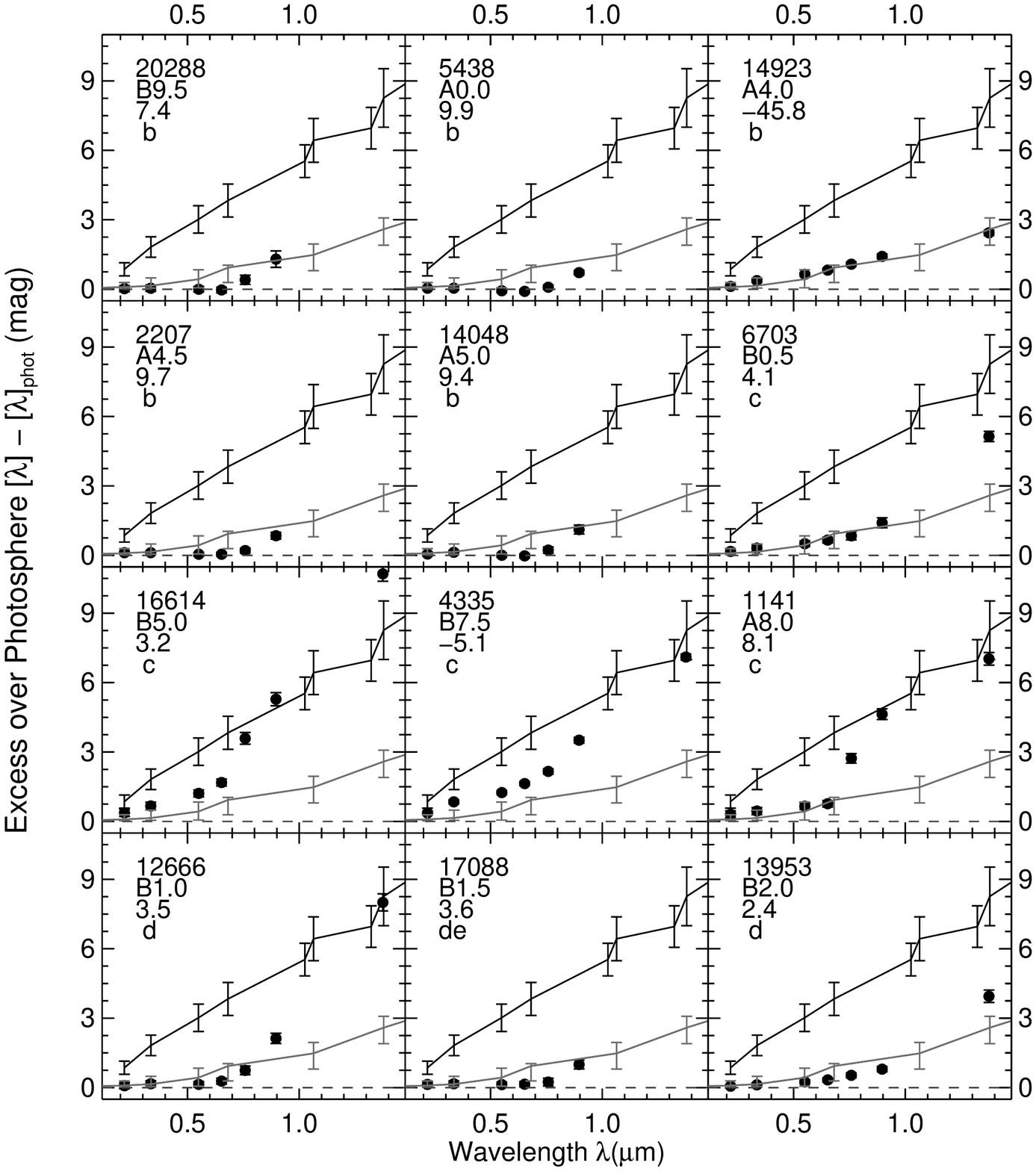,width=0.99\linewidth,clip=}
\end{center}
\end{figure}

\begin{figure}
\begin{center}
\epsfig{file=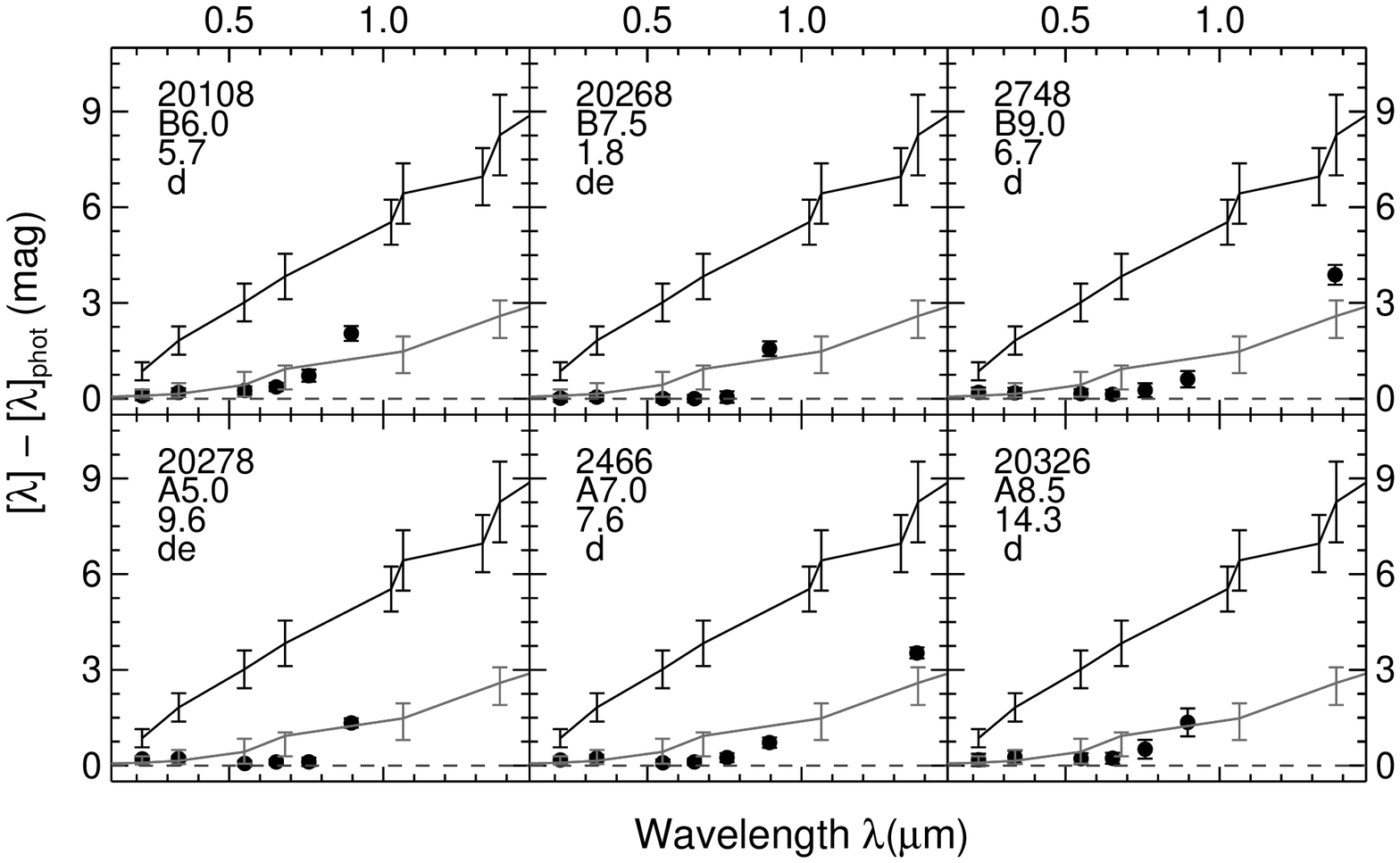,width=0.99\linewidth,clip=}
\end{center}
\end{figure}

\end{document}